\documentclass[preprint2,twoside]{hwo}

\usepackage{graphicx}
\usepackage[version=4]{mhchem}
\usepackage{multirow}
\usepackage[table,xcdraw]{xcolor}
\usepackage{tabularx}
\usepackage{tabularray}

\defcitealias{Astro2020}{Astro2020}

\input{hwo.h}

\begin{document}

\title{\textbf{\LARGE Habitable Worlds Observatory (HWO): Living Worlds Working Group: The Search for Life on Potentially Habitable Exoplanets}}
\author {\textbf{\large {\textit{Co-Chairs:} Giada Arney$^{1}$, Niki Parenteau$^{2}$; \textit{Sub-Group Co-Chairs:} Natalie Hinkel$^{3}$, Eric Mamajek$^{4}$, Joshua Krissansen-Totton$^{5}$, Stephanie Olson$^{6}$, Edward Schwieterman$^{7}$, Sara Walker$^{8}$; \textit{Steering Committee:} Kevin Fogarty$^{2}$, Ravi Kopparapu$^{1}$, Jacob Lustig-Yaeger$^{9}$, Mark Moussa$^{1}$, Sukrit Ranjan$^{10}$, Garima Singh$^{11}$, Clara Sousa-Silva$^{12}$ }}}

\affil{$^1$\small\it NASA Goddard Space Flight Center, Greenbelt, MD, USA} 
\email{giada.n.arney@nasa.gov}
\affil{$^2$\small\it NASA Ames Research Center, Moffett Field, CA, USA}

\affil{$^3$\small\it Louisiana State University, Baton Rouge, LA, USA}
\affil{$^4$\small\it Jet Propulsion Laboratory, California Institute of Technology, Pasadena, CA, USA}
\affil{$^5$\small\it University of Washington, Seattle, WA, USA}
\affil{$^6$\small\it Purdue University, West Lafayette, IN, USA}
\affil{$^7$\small\it University of California at Riverside, Riverside, CA, USA}
\affil{$^8$\small\it Arizona State University, Tempe, AZ, USA}
\affil{$^9$\small\it Johns Hopkins University Applied Physics Laboratory, Baltimore, MD, USA}
\affil{$^{10}$\small\it University of Arizona Lunar and Planetary Laboratory, Tucson, AZ, USA}
\affil{$^{11}$\small\it Gemini North Observatory, Hilo, HI, USA}
\affil{$^{12}$\small\it Bard College, Red Hook, NY, USA}

\author{\small{\bf Contributing Authors:} Ruslan Belikov (NASA Ames Research Center), Maxwell Frissell (University of Washington), Samantha Gilbert-Janziek (University of Washington), Vincent Kofman (University of Oslo), Natasha Latouf (NASA Goddard Space Flight Center), Mary Anne Limbach (Jet Propulsion Laboratory), Rhonda Morgan (Jet Propulsion Laboratory), Christopher Stark (NASA Goddard Space Flight Center), Armen Tokadjian (Jet Propulsion Laboratory), Anna Grace Ulses (University of Washington), Nicholas Wogan (NASA Ames Research Center), Mike Wong (Carnegie Science), Amber Young (NASA Headquarters)}

\author{\footnotesize{\bf Endorsed by:}
Eleonora Alei (NASA Goddard Space Flight Center, Natalie Allen (Johns Hopkins University), Narsireddy Anugu (Georgia State University), David Arnot (The Open University), Reza Ashtari (Johns Hopkins University-Applied Physics Laboratory), Amedeo Balbi (Università di Roma Tor Vergata, Italy), Komal Bali (ETH Zurich), Sarah Barbosa (Universidade Federal do Ceara), Jacob Bean (University of Chicago), Ruslan Belikov (NASA Ames Research Center), Katherine Bennett (Johns Hopkins University), Alan Boss (Carnegie Science), Kara Brugman (Arizona State University), Zachary Burr (ETH Zurich), José A.Caballero (Centro de Astrobiología CSIC-INTA), Douglas Caldwell (SETI Institute), Oliver Carey (Brown University), Aarynn Carter (Space Telescope Science Institute), Jessie Christiansen (Caltech/IPAC), Ligia F Coelho (Cornell University), Giuseppe Conzo (Gruppo Astrofili Palidoro), Jaime Crouse (Johns Hopkins University/NASA Goddard Space Flight Center), Nicolas Crouzet (Kapteyn Astronomical Institute, University of Groningen, The Netherlands), Ruben Joaquin Diaz (NSF NOIRLab), Jamie Dietrich (Arizona State University), Steven Dillmann(University), Luca Fossati (Space Research Institute, Austrian Academy of Sciences), Kevin France (University of Colorado), Megan Gialluca (University of Washington), Ana Ines Gomez de Castro (Universidad Complutense de Madrid), Darío González Picos (Leiden Observatory), Kenneth Goodis Gordon (University of Central Florida), Olivier Guyon (University of Arizona, NAOJ and ABC), Caleb Harada (UC Berkeley), Sonny Harman (NASA Ames Research Center), Xinchuan Huang (SETI Institute \& NASA Ames Research Center), Chris Impey (University of Arizona), Markus Janson (Stockholm University), Berger Jean-Philippe (IPAG, Université Grenoble Alpes, CNRS), Renaud Joe (University of Maryland / NASA Goddard), Jens Kammerer (European Southern Observatory), Theodora Karalidi (University of Central Florida), Preethi Karpoor (Indian Institute of Astrophysics), James Kasting (Penn State University), Finnegan Keller (Arizona State University), Alen Kuriakose (KU Leuven, Belgium), Adam Langeveld (Johns Hopkins University), Eunjeong Lee (EisKosmos (CROASAEN), Inc.), Mercedes López-Morales (Space Telescope Science Institute), Evelyn Macdonald (University of Vienna), Jack Madden (Blue Marble Space Institute of Science), Elena Manjavacas (Space Telescope Science Institute), Melinda Soares-Furtado (UW-Madison), Drew Miles (California Institute of Technology), David Montes (UCM, Universidad Complutense de Madrid), Faraz Nasir Saleem (Egyptian Space Agency (EgSA)), Eric Nielsen (Department of Astronomy, New Mexico State University), Sarah Peacock (UMBC/NASA Goddard Space Flight Center), Andreas Quirrenbach (Landessternwarte, U Heidelberg), Patricio Reller (University College London), Ignasi Ribas (Institute of Space Sciences (ICE, CSIC) \& Institute of Space Studies of Catalonia (IEEC)), William Roberson (New Mexico State University), Blair Russell (Chapman University), Farid Salama (NASA Ames Research Center), Arnaud Salvador (German Aerospace Center (DLR)), Aniket Sanghi (Caltech), Gaetano Scandariato (INAF), Everett Schlawin (University of Arizona), Laura Silva (INAF-OATs, Italy), Christopher Stark (NASA Goddard Space Flight Center), Sarah Steiger (Space Telescope Science Institute), Tomas Stolker (Leiden University), Johanna Teske (Carnegie Science), Thaddeus Komacek (University of Oxford), Christohper Theissen (UC San Diego), Armen Tokadjian (Jet Propulsion Laboratory), Martin Turbet (LMD, LAB, IPSL, CNRS), MaggieBeth Turcotte (NASA Goddard Space Flight Center (CRESST/SURA)), Vincent Van Eylen (UCL), Connor Vancil (UCSB), Mariela C.Vieytes (Instituto de Astronomía y Física del Espacio (IAFE, CONICET-UBA) Argentina), Iva Vilović (Leibniz Institute for Astrophysics Potsdam (AIP)), Austin Ware (Arizona State University), Jessica Weber (Jet Propulsion Laboratory), Megan Weiner Mansfield (University of Maryland)
}


  
\section{Abstract} 

  The discovery of a biosphere on another planet would transform how we view ourselves, and our planet Earth, in relation to the rest of the cosmos. We now know Earth is one planet among eight circling our sun; our sun is part of a swirling galaxy of over one hundred billion other suns; and our galaxy is one of untold billions in the universe. While we do not yet know how many – if any – other biospheres exist on the countless worlds orbiting countless other suns, we stand at the precipice of a new era of discovery, enabled by powerful new facilities able to peer across the light years into the atmospheres of planets similar to our own. 
 \textit{This article is an adaptation of a science case document (SCDD) developed for the NASA Astrophysics Flagship mission the Habitable Worlds Observatory (HWO) Science, Technology, and Architecture Review Team (START) Living Worlds Working Group.}

\section{Science Goal}
\textbf{Are Earth-like global biospheres common or rare in the galaxy? }

\textit{“Inspired by the vision of searching for signatures of life on planets outside our solar system…the priority recommendation in the frontier category for space is a large IR/O/UV telescope…”  - \citetalias{Astro2020}
}

In recent years, astronomers have constrained the frequency of rocky planets orbiting in their stars’ habitable zones – the region around stars where “Earth-like” conditions and surface liquid water are possible. It is estimated that about one fifth of all stars in our galaxy may have a rocky planet in their habitable zones. Yet, despite knowing the general frequency of planets that might be Earth-like, we do not yet know how many of these worlds actually host habitable conditions. Further, we can only speculate about the fraction of habitable planets that actually host life. The Habitable Worlds Observatory (HWO) will be the first facility ever designed to address the millennia-old question “Are we alone?”  By placing the first constraints on the frequency of habitable – and inhabited – planets, HWO’s large aperture and robust instrument suite may help us understand just how alone we are – or discover our universe is teeming with other living worlds. 

\textit{“If planets like Earth are rare, our own world becomes even more precious. If we do discover the signature of life on another planetary system, it will change our place in the universe in a way not seen since the days of Copernicus.”  - \citetalias{Astro2020}
}

\section{Science Objective}
\textbf{Discover biosignatures on exoplanets in a false positive/false negative framework, or discover statistically meaningful limits for null results. 
}

\begin{figure*}[ht!]
    \centering
    \includegraphics[width=0.85\textwidth]{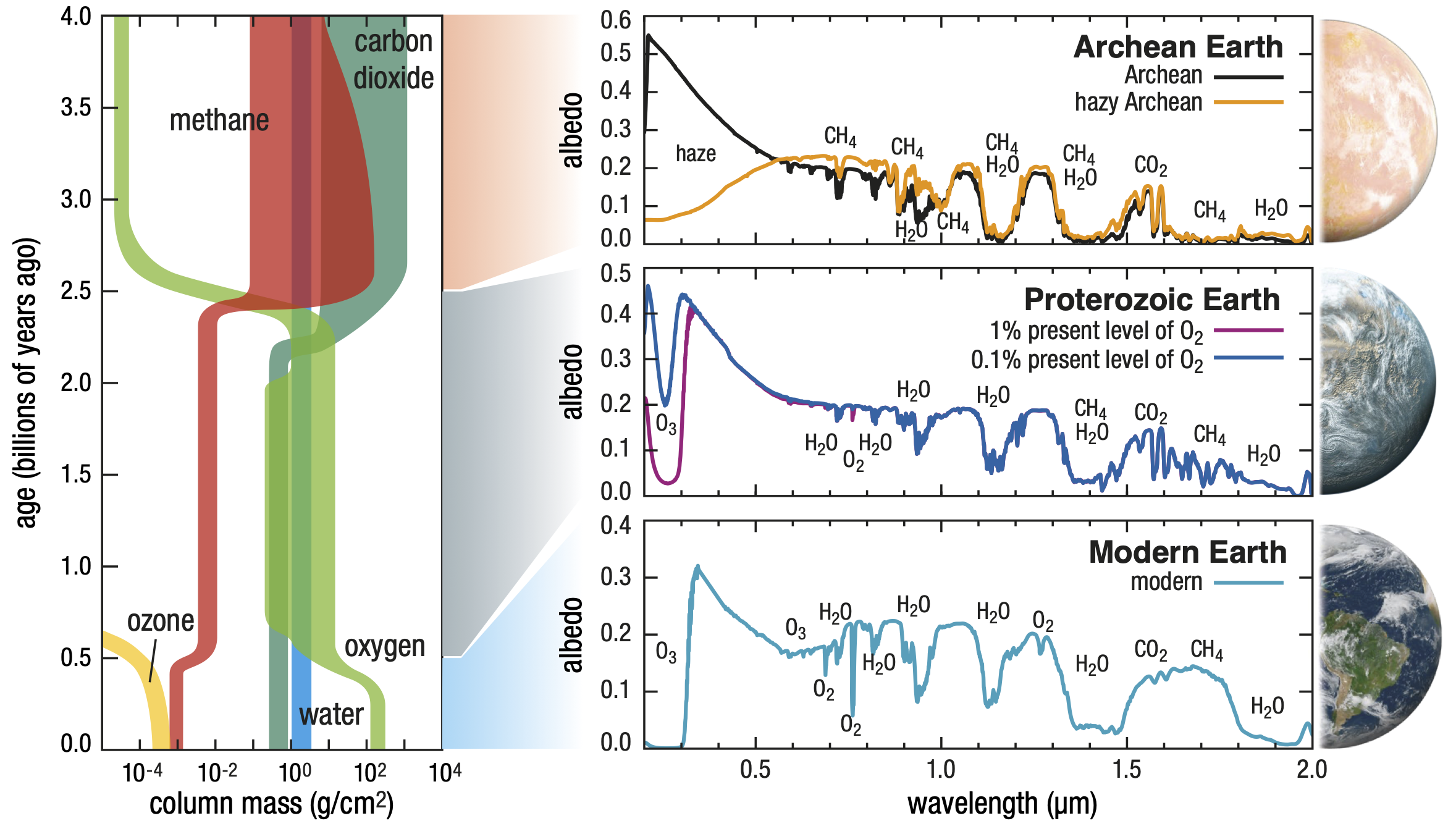}
    \caption{Biosignatures and habitability indicators of Earth through time reflect how life on our planet has co-evolved with its environment. The biosignatures of Earth through time are a useful minimum set of spectral features to seek on exoplanets with HWO.}
    \label{fig:fig1}
\end{figure*}

HWO will offer an unprecedented platform to search for remotely observable signs of life, termed “biosignatures,” in the spectra of exoplanets. Life impacts our planet in innumerable ways. Astrobiologists frequently speak of life as a planetary process – one that both modifies and is modified by its environment (Fig.~\ref{fig:fig1}). Yet when observing planetary points of light from across the light years, only a small subset of countless life’s impacts on its planet will be remotely observable: those that result from planet-wide, global processes. 

Every potential biosignature that we detect must be carefully considered in the context of its environment to mitigate against being confused by abiotic “false positive” mimics. The amount of any gas that we can measure in a planetary atmosphere is the result of competition between source and sink processes. We must obtain enough information about the environment to determine whether a given gas is present because it has a high production rate in an environment with strong sinks, or whether it can be explained by a modest, abiotic production rate in an environment with smaller sinks. These considerations were outlined in the 2022 Biosignatures Standards of Evidence (SoE) Workshop and white paper \citep{Meadows2022} wherein the authors presented a generalized framework consisting of five scientific questions that would guide the evaluation of a claim of life detection corresponding to both biosignature detection and interpretation. The Living Worlds SCDD is designed to ensure that HWO has the capabilities to follow each step outlined in this standards of evidence framework. 

\begin{figure*}[ht!]
    \centering
    \includegraphics[width=0.85\textwidth]{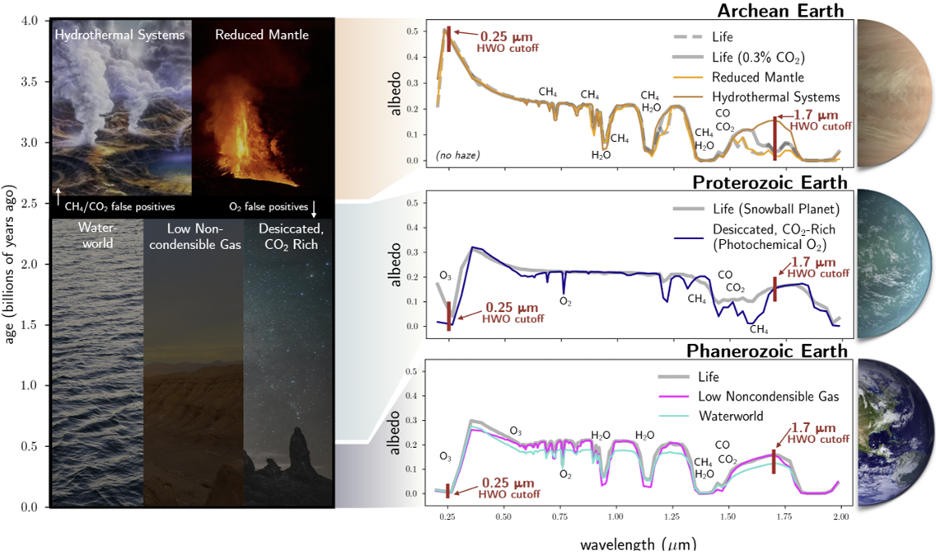}
    \caption{Biosignatures must be considered in the context of their environments to rule out false positives. Possible false positive planets are shown, with relevant spectral features labeled. False positives relevant to oxygenated planets are on the bottom (Phanerozoic, or modern, Earth and Proterozoic Earth); those relevant to anoxic planets (Archean Earth) are on the top. Figure by Samantha Gilbert-Janizek (UW).}
    \label{fig:fig2}
\end{figure*}

Astrobiologists frequently speak of looking for atmospheres in chemical disequilibrium when seeking biosignatures on exoplanets. Chemical disequilibrium is a state in which two or more chemical species that are incompatible coexist, and it can be quantified by the available free energy in the system \citep[e.g.,][]{krissansen2016detecting, wogan2020chemical}. Oxygen and methane are examples of gases that coexist in chemical disequilibrium in modern Earth’s atmosphere. Oxygen radicals rapidly destroy methane – its lifetime in Earth’s atmosphere is only about 10 years – so robust continual production is required to explain its presence. Observing chemical disequilibrium on an exoplanet can suggest life \citep[e.g.,][]{hitchcock1967life}, because disequilibria can be challenging to maintain through slow, abiotic production rates of gases –  although exceptions exist \citep{wogan2020chemical} , so planetary context must be considered to rule out false positives (Fig.~\ref{fig:fig2}).

\begin{figure*}[ht!]
    \centering
    \includegraphics[width=0.85\textwidth]{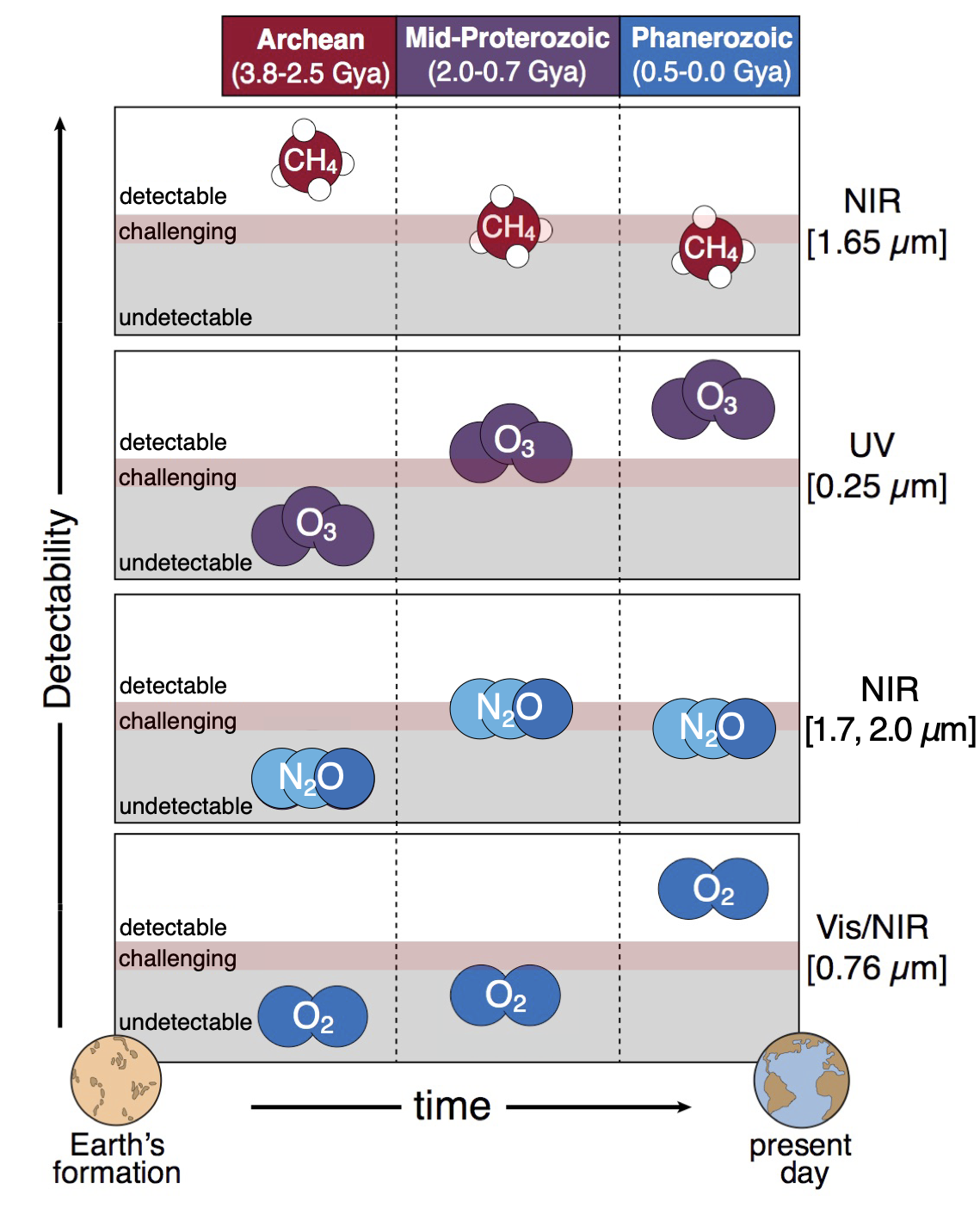}
    \caption{The dominant biosignatures of Earth, and their detectability, have varied over our planet’s geological history. Different photochemistries around different stars can also alter the relative detectability of biosignatures. A wavelength range from the NUV to the NIR ensures that the dominant biosignatures over Earth history can be detected on exoplanets. }
    \label{fig:fig3}
\end{figure*}

Given that our planet is our only example of a habitable world with life, strategies to search for life on exoplanets generally start with considerations of how to detect life on Earth \citep[e.g.,][]{sagan1993search}. Yet our planet has not always looked the way it does today (Fig.~\ref{fig:fig1}). In the same way that peering across interstellar distances will reveal diverse exoplanets unlike our own, looking back through geological time shows us different types of dominant biospheres and environments that prevailed on our world over Earth’s history. Thus, if we want to search for signs of life on “Earth-like” exoplanets, we should, at minimum, design an observatory capable of detecting signs of life on exoplanets similar to the different faces of Earth over its history. The dominant biosignatures of our planet have varied considerably over Earth history (Fig.~\ref{fig:fig3}), requiring a wide wavelength range spanning from the NUV to the NIR to characterize each inhabited phase of our planet over geological time. An observatory not capable of detecting the dominant, planet-wide spectral features indicative of habitability and life on planets analogous to Earth through time will run the risk of being under-designed to meet its eponymous central goal. 

Briefly, Earth’s major geological periods encompass the following. 

The earliest eon of Earth history, the \textbf{Hadean} (before 4 billion years ago) is very poorly constrained due to an extremely sparse geological record. Yet Earth was likely habitable by the end of the Hadean \citep{Harrisonetal2005}, and there is evidence that life may have arisen during this period \citep[e.g.,][]{Belletal2015,Nutmanetal2016}. Large impacts may have induced transient highly reducing atmospheric states \citep{Zahnleetal2010}, which may have been important for prebiotic chemistry. Photochemical and lightning reactions in the Hadean atmosphere may also have contributed chemical precursors to life \citep[e.g.,][]{Woganetal2023} and metabolically useful molecules \citep{Kasting2014}. 

The \textbf{Archean} (4-2.5 billion years ago) boasted an anoxic atmosphere and a thriving microbial, anaerobic biosphere. It is likely that Archean Earth’s atmosphere was rich in greenhouse gases (e.g., carbon dioxide, \ce{CO2} and methane, \ce{CH4}) to warm our planet to hospitable temperatures under the illumination of our fainter sun at this time period (75$\%$-80$\%$ the modern solar luminosity). Atmospheric methane abundances, in particular, may have been 2-3 orders of magnitude greater than today \citep{Pavlovetal2000} due to robust production of this gas by methane-producing methanogen microorganisms \citep{WoeseFox1977, Uenoetal2006} and a long atmospheric lifetime in an oxygen-poor environment. However, methane can also be produced by abiotic processes, so care must be taken when interpreting this gas as a biosignature \citep[e.g.,][]{Krissansen2022understanding}. The abundances of other carbon-bearing species are important pieces of contextual information required to interpret methane as a biosignature in anoxic atmospheres: specifically, the abundances of \ce{CH4}, \ce{CO2}, and CO must be well-constrained, as must the photochemical context generated by the UV spectrum of the host star \citep{Thompson2022case, Schwieterman2019_CO}. 

For planets with high amounts of \ce{CH4} relative to \ce{CO2} (e.g. \ce{CH4}/\ce{CO2} ratios in excess of $\sim$0.2), organic haze particles can form \citep[e.g.,][]{Trainer2006organic}, dramatically alternating the planet’s spectral appearance at UV-blue wavelengths, its climate (given that hazes have an anti-greenhouse effect), and its surface habitability (given that hazes are strong UV absorbers) \citep{Arneyetal2016, Arneyetal2017, Arneyetal2018}. Organic haze may have transiently existed in Earth’s atmosphere during the Archean \citep[e.g.,][]{Zerkle2012bistable}, challenging the notion that all “Earths” will be the “pale blue dots.” Hazy Archean Earth, appearing a pale orange to remote observers, is arguably the most “alien” planet for which we have geochemical data, reminding us of the importance of designing an observatory capable of characterizing planets with a wide variety of properties dissimilar to modern Earth. 

The \textbf{Proterozoic} eon (2.5 billion years ago – 541 million years ago) began with the marked rise of oxygen (\ce{O2}) in our planet’s atmosphere, irreversibly altering our planet’s remotely observable characteristics in numerous ways. Additionally, the rise of oxygen and its photochemical byproduct, ozone (\ce{O3}), provided a powerful ultraviolet shield that may have eased the transition of life from the seas onto land. Yet during the mid-Proterozoic, oxygen levels may have been as low as 0.1$\%$ of the present atmospheric level \citep{Planavsky2014low, Lyons2014rise, Lyons2021oxygenation}. Methane atmospheric abundances may likewise have been suppressed to low levels, possibly low enough to render methane’s spectral features largely invisible to an observatory like HWO \citep{Olson2016limited}, though the strongest band at 1.65 $\mu$m could be observable for \ce{CH4} abundances modestly higher than modern Earth’s (e.g., $\sim$10 ppm compared to modern Earth’s $\sim$1.6 ppm). An additional biosignature that could be sought on Proterozoic-like exoplanets is \ce{N2O} \citep[e.g.,][]{Buick2007, Schwieterman2022}, although \ce{N2O}’s relatively weak spectral features $<$ 2 $\mu$m and these features’ overlap with \ce{H2O} features suggest that only high \ce{N2O} abundances ($\sim$100-1,000 ppm) might be detected \citep{tokadjian2024detectability}. 

Based on what is known about the mid-Proterozoic, a strong spectral feature from \ce{O3} in the near UV (NUV) is the best – and possibly only – remotely observable sign of life for a facility like HWO. The best opportunity to capture a disequilibrium biosignature (e.g., between \ce{O3} and \ce{CH4}, or \ce{O3} and \ce{N2O}) requires wavelength coverage from $\sim$0.25 $\mu$m to $\sim$1.7 $\mu$m to fully capture the \ce{O3} Hartley band ($\sim$0.3 $\mu$m cutoff), and strong 1.65 $\mu$m \ce{CH4} band in its entirety, and possibly the strongest NIR \ce{N2O} bands (1.52, 1.68, 1.78 $\mu$m). It cannot be overstated: the mid-Proterozoic, which lasted for about a billion years, represents a planetary archetype that would pose a formidable challenge for life-detection on a planet analogous to our own. An observatory under-designed to meet this challenge runs the risk of a false negative detection of life \citep[e.g.,][]{Reinhard2017} on similar exoplanets. False negative detections of life on weakly oxygenated planets similar to mid-Proterozoic Earth may pose a greater obstacle to searches for life beyond the Earth than false positive detections of life on uninhabited planets: given that oxygen is a highly reactive gas, there are numerous barriers to its accumulation in planetary atmospheres even if a biosphere is producing it \citep[e.g.,][]{Lyons2014rise}. 

\textbf{Modern, or Phanerozoic Earth} (541 million years ago – present) has been scrutinized as an analog for exoplanets \citep[e.g.,][]{sagan1993search,robinson2018earth}. Modern Earth is and always will be our best-studied, best understood world, providing invaluable information to inform future mission designs. Important gases in modern Earth’s atmosphere include \ce{O2, O3, CH4, and CO2}. While the simultaneous presence of oxygen and methane have been pointed to as an important gas pair indicative of an atmosphere in chemical disequilibrium that is extremely difficult or even impossible to explain without life \citep[e.g.,][]{hitchcock1967life}, methane may be challenging to detect in the atmosphere of a modern Earth twin due to precisely the same process that renders it a valuable biosignature: oxygen radicals sourced from ozone photochemistry destroy methane rapidly, so its photochemical lifetime in the atmosphere is only about 10 years, and it is present at a scant part-per-million abundance with minor spectral impact in the HWO wavelength range. In absence of a methane detection, the only way to rule out the main false positive process that can produce oxygen around sun-like stars is to constrain other atmospheric constituents, including background gases, total atmospheric pressure, and surface composition/climate (see below). However, different photochemical consequences in the atmospheres of planets around other types of stars add nuance to this picture, and it may be easier to detect oxygen and methane together for modern Earth-like planets orbiting stars of lower mass than the sun \citep[e.g.,][]{segura2005MDwarfs, arney2019k}.

It has been suggested that our planet entered a new geological epoch for the most recent sliver of Earth history: the \textbf{Anthropocene}, a period of significant human impact on the Earth, with various proposed start dates ranging from the Neolithic period to as recently as the last decades. While this period is not officially recognized at present, considering it reminds us that a different type of biosignature, technosignatures, might be possible to remotely detect on exoplanets. While biosignatures are signs of life in general, technosignatures are remotely observable signs of technology. 


The reasons for the focus on Earth through time are not purely Earth-centric. Methane-producing metabolisms are relatively simple compared to other biogenic gas-producing metabolisms, and it's reasonable to speculate that they may evolve on other worlds given the probable ubiquity of the necessary substrates (\ce{H2}, \ce{CO2}). Indeed, methanogenesis emerged early in Earth’s evolution \citep{weiss2016physiology, wolfe2018horizontal}. Similarly, oxygen biosignatures from photosynthesis may also be ubiquitous since the substrates (\ce{H2O}, \ce{CO2}) would be widely available on a habitable exoplanet. Compared to geochemical electron donors such as \ce{H2}, which are typically in low supply in limited environments, the distribution of \ce{H2O} that serves as an electron donor for oxygenic photosynthesis would allow for global-scale proliferation in surface environments. However, it must be noted that the 'metabolic machinery' of the photosynthetic subsystems is complex and it's not a foregone conclusion that oxygenic photosynthesis would evolve on exoplanets. 

With that said, it would be more surprising to discover that all living worlds look like Earth through time than it would be to discover worlds radically different from Earth. Therefore, while we should ensure that HWO is capable of sensing signs of life on planets like Earth, it is also important to design an observatory with sufficiently robust capabilities that maximize our chances of observing and correctly interpreting biosignatures on planets with different histories and alien (in every sense of the word) biospheres. The history of exoplanet discoveries reminds us that we must be prepared for the unexpected as we consider the existence of exoplanets not represented by solar system analogs (e.g., Hot Jupiters, mini Neptunes, super Earths) and discovery of unexpected processes (e.g., planetary migration). There is no reason to think that our first observations of potentially habitable and inhabited planets will be any less surprising. If anything, we should expect the challenges of interpreting living worlds to be significantly greater than the challenges of interpreting other kinds of processes, especially given our singular world for comparison. Meeting this challenge demands an observatory with the most robust capabilities possible (e.g., wide wavelength range, large aperture, small inner working angle, etc).  

\begin{figure*}[ht!]
    \centering
    \includegraphics[width=0.85\textwidth]{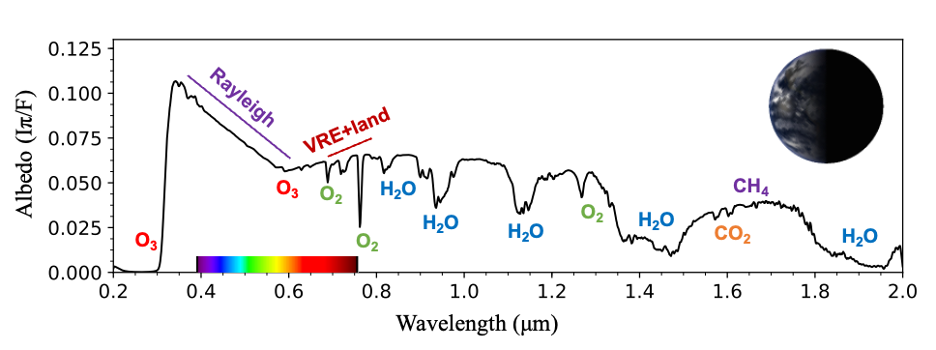}
    \caption{Spectral features of Modern Earth at quadrature phase (half illumination). This figure is reproduced with minor modifications from Schwieterman et al. (2018a) under Creative Commons Attribution License CC-BY 4.0. Figure modifications include labels of additional gas features, an inset of a simulated Earth at half illumination, and an inset color bar indicating the visible wavelength range.
 }
    \label{fig:fig4}
\end{figure*}

There are numerous biosignature gases to be sought \citep[e.g.,][]{seager2012astrophysical, kaltenegger2017characterize, schwieterman2018exoplanet, catlingdavid2018exoplanet}, as well as other planetary features that will help us interpret those features in the context of their environments. A useful Objective to bound HWO capabilities is capturing ozone in the NUV and methane, carbon dioxide, and carbon monoxide in the NIR near 1.7 $\mu$m, for as many targets as possible. In between these wavelengths, other key gases can be observed: \ce{O2}, \ce{H2O}, \ce{CO2}, \ce{O2-O2}, and others (Fig.~\ref{fig:fig1}-~\ref{fig:fig4}). Here, we describe key categories of biosignature gases and other spectral features required to understand those detections in the proper context. 

\subsection{Biosignatures in Oxidized Atmospheres}

Molecular oxygen (\ce{O2}) has generally been considered the highest priority biosignature gas to seek in exoplanet atmospheres \citep[e.g.,][]{meadows2017reflections}. The byproduct of our planet’s dominant metabolism, oxygenic photosynthesis, \ce{O2} has profoundly changed our atmosphere – and spectrum – over Earth’s geological history. While other more primitive forms of anoxygenic photosynthesis are possible, they are limited by availability of reductants (e.g., \ce{H2}, \ce{H2S}, \ce{Fe2+}) and would only be detectable via surface biosignatures of their photosynthetic pigments. Oxygenic photosynthesis, meanwhile, uses cosmically ubiquitous compounds: \ce{H2O}, starlight (h$\nu$), as well as \ce{CO2}, which is expected to occur generally on terrestrial planets:

\ce{CO2 + 2H2O + h$\nu$ -> (CH2O)$_{org}$ + H2O + O2}

Ozone is a photochemical byproduct of atmospheric \ce{O2}. The reactions that produce and destroy \ce{O3} on Earth are generally called the Chapman reactions \citep{chapman1930}:

\ce{O2 + h$\nu$ -> 2O}

\ce{O + O2 + M -> O3 + M} 

\ce{O3 + h$\nu$ -> O2 + O}

\ce{O + O3 -> 2O2} 

Here, M is an inert molecule or neutral atom that can carry away excess energy and momentum. Naturally, given that these reactions depend on photochemistry, different stars may be more or less efficient at generating – or destroying – ozone \citep[e.g.,][]{segura2003ozone, grenfell2014sensitivity, arney2019k}. Additionally, energetic particles from stellar activity can deplete \ce{O3} \citep{tabataba2016atmospheric, tilley2019modeling}. In a general sense, understanding the photochemical context of an atmosphere is critical for evaluating biosignatures.

As described above, ozone is a critical indicator of weakly oxygenated planets such as Proterozoic Earth, because it has a strong spectral feature at NUV wavelengths even for planets with little \ce{O2}. This feature may be the only observable biosignature at HWO wavelengths for planets similar to the mid-Proterozoic, so access to it is critical for guarding against false negative detections of life on weakly oxygenated planets with low amounts of atmospheric \ce{CH4}. Additionally, changes in seasonal oxygen production on weakly oxygenated planets might be detected by modulations in the strength of NUV ozone feature as a type of temporal biosignature \citep{olson2018atmospheric}. 

\subsubsection{Oxygen biosignature false positives}

\begin{figure*}[ht!]
    \centering
    \includegraphics[width=0.85\textwidth]{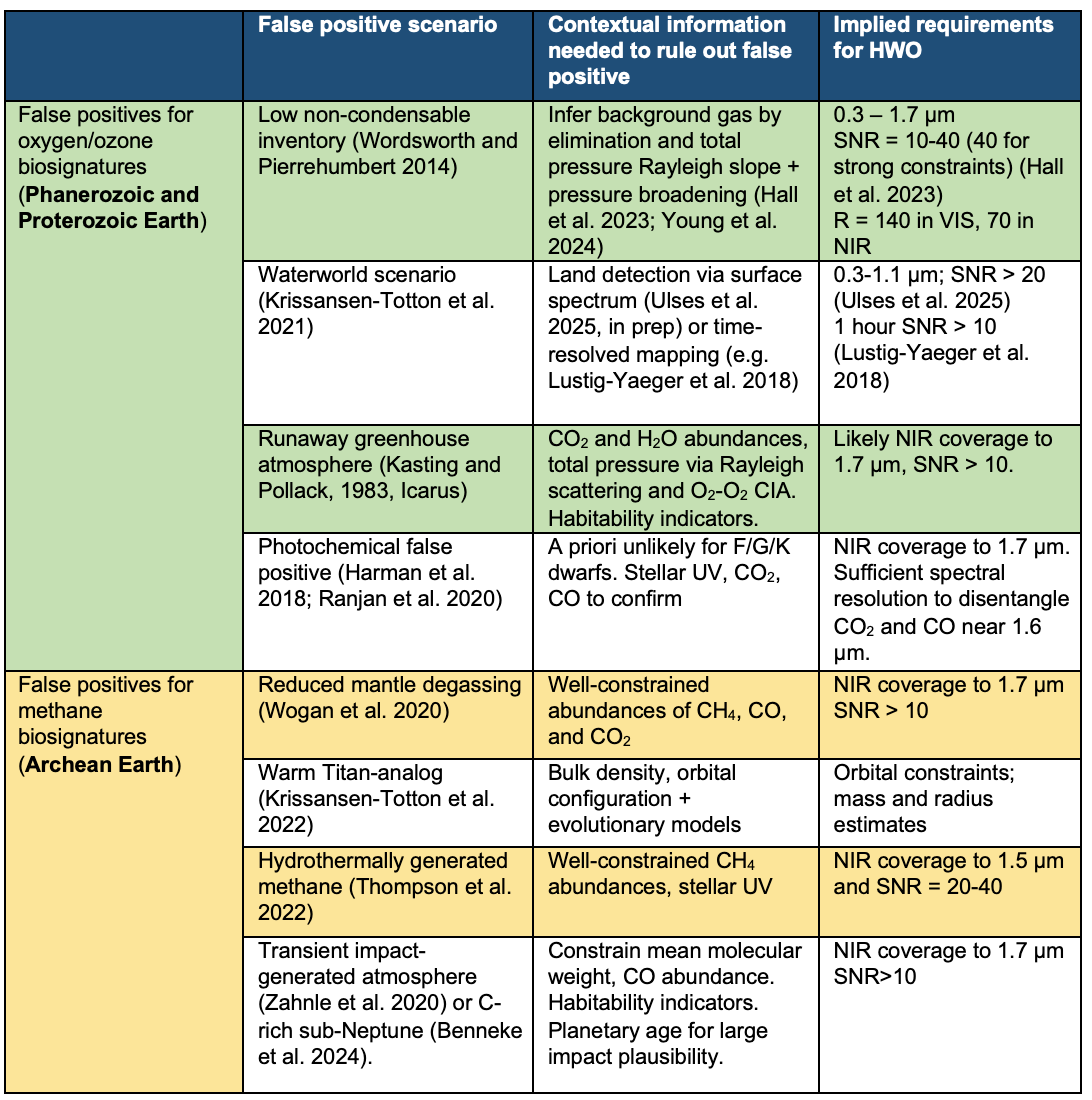}
    \caption{Types of biosignatures and false positives for different types of environments. 
 }
    \label{fig:fig5}
\end{figure*}

Detection of \ce{O2} or \ce{O3} in an exoplanet atmosphere would be a watershed moment, but this detection alone would not prove that there is life on the planet, for there are abiotic processes that can produce an atmosphere with detectable \ce{O2} that must be considered \citep{meadows2017reflections}. Fig.~\ref{fig:fig5} summarizes information on various oxygen false positive scenarios. Most of these false positive scenarios primarily apply to M dwarf stars and could result in massive quantities (i.e., bars) of oxygen through water-loss processes. Such oxygen-rich atmospheres could be identified through detection of \ce{O2-O2} collisional induced absorption features\citep{schwieterman2016identifying}. More relevant to the sun-like (F/G/K) stars that represent the majority of the HWO target stars is false positive oxygen generated by the following mechanisms:

\begin{itemize}
\item On planets with low non-condensable gas inventories (e.g., \ce{N2}), water vapor can more easily reach the upper atmosphere \citep{wordsworthpierrehumbert2014} where it can be photolyzed to generate false positive O2, followed by loss to space of the lightweight liberated H. Background non-condensable gases can be constrained by observing a planet’s Rayleigh scattering slope, as well as through observations of pressure broadening of spectral features \citep{young2024retrievals}. The background gas abundance and composition may also be inferred by elimination of other plausible background gas constituents \citep{Hall2023constraining}. However, access to NIR wavelengths ($\sim$1.7 $\mu$m) is necessary to distinguish between a \ce{N2} background and CO background, since the latter could indicate a photochemical oxygen biosignature false positive.

\item Planets with extremely deep surface oceans may accumulate significant abiotic oxygen via H escape since the pressure overburden of $\sim$50+ Earth oceans suppresses oxygen sinks \citep{krissansen2021oxygen}. Such a scenario can be ruled out by land detection, which puts an upper limit on ocean depth. Land detection requires robust surface characterization capabilities.

\item Planets in a moist or runaway greenhouse state can experience \ce{O2} accumulation from elevated H loss following \ce{H2O} photolysis. Such planets are likely to occur closer to their stars than the conservative habitable zone boundaries \citep[e.g.,][]{kopparapu2013habitable}. Constraints on orbit as well as \ce{CO2} and \ce{H2O} abundance will be required to rule out these false positives. 

\item Planets with high \ce{CO2} abundances and low abundances of hydrogen-bearing molecules can experience \ce{O2} buildup due to the breakdown of the OH-mediated catalytic cycles that recombine photolyzed \ce{CO2} \citep{gao2015stability}. This mechanism would also accumulate abundant \ce{CO}, so upper limits on CO abundance can rule out this false positive. Additionally, the detection of hydrogen-bearing species in the planet’s atmosphere can rule out this false positive. 
\end{itemize}


Note that many of the photochemical false positives for oxygen have recently been eliminated by model intercomparisons, reducing the ambiguity of oxygen as a biosignature \citep{Ranjan2020Photochemistry,ranjan2023importance}. Note that all of these oxygen false positive scenarios would be strongly disfavored by a simultaneous detection of atmospheric methane. This is because the implied kinetic disequilibrium is extremely challenging to explain without large methane fluxes, which would not be expected for any of the scenarios above, or indeed for any plausible abiotic process. While it is challenging to detect oxygen and methane simultaneously in the photochemical context of modern Earth orbiting the sun, the photochemical lifetime of methane is longer around lower mass stars such as M dwarfs \citep{segura2005MDwarfs} and K dwarfs \citep{arney2019k}, so these stars likely offer better prospects for simultaneously detecting this key gaseous disequilibrium pair.

\subsection{Biosignatures in Anoxic Atmospheres}

For anoxic atmospheres like Archean Earth, atmospheric methane is the most promising biosignature candidate. Archean Earth likely had similar rates of methane production to the modern Earth \citep{Kharecha2005, Sauterey2020}, but this gas’s longer photochemical lifetime in an anoxic atmosphere would have enabled orders of magnitude more methane to accumulate in the atmosphere of Archean Earth compared to modern Earth, enhancing its detectability. 

\subsubsection{Methane biosignature false positives}
With that said, methane\textemdash{}just as with oxygen\textemdash{}can be produced by abiotic processes, so careful consideration of potential “false positives” is needed before a biological source can be inferred. 

\begin{figure*}[ht!]
    \centering
    \includegraphics[width=0.85\textwidth]{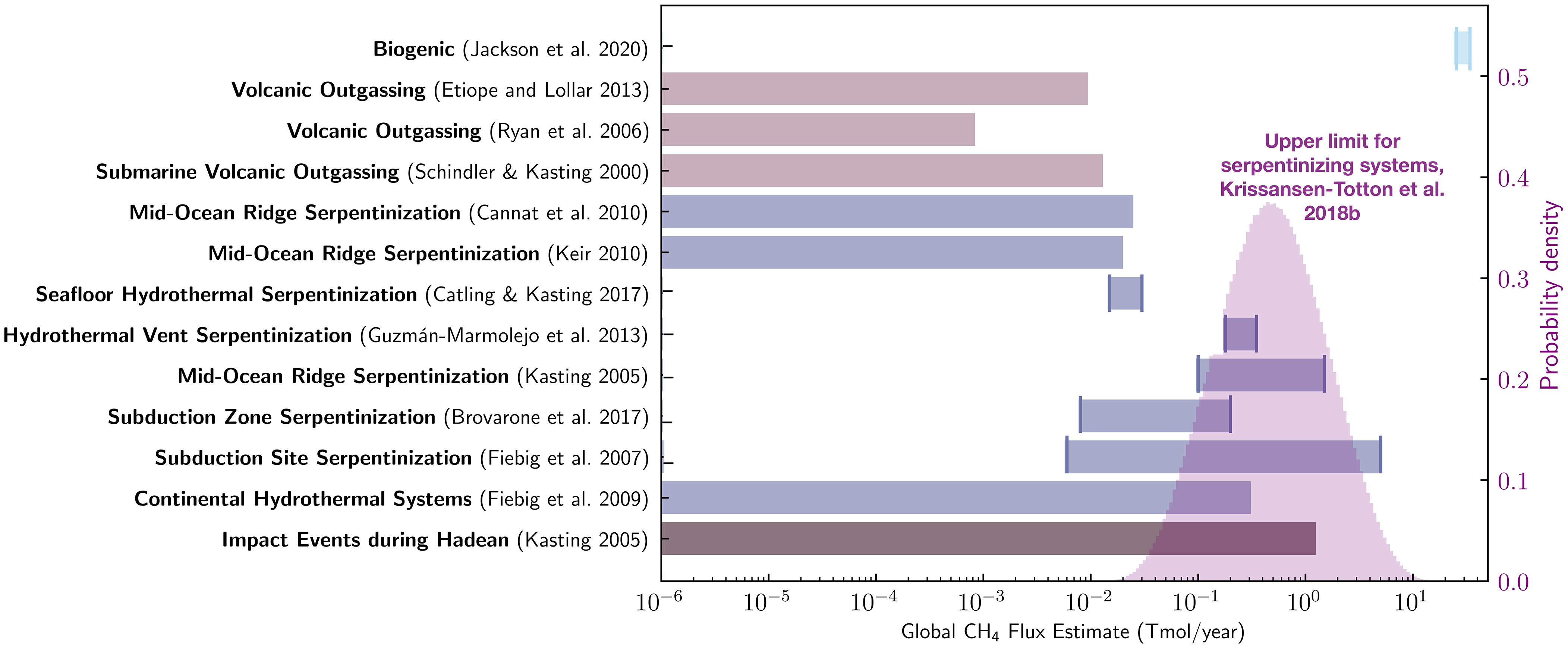}
    \caption{Abiotic sources of methane compared to Earth’s modern biogenic flux, which is more rapid than all abiotic sources. Reproduced with permission from Thompson et al. (2022).
 }
    \label{fig:fig6}
\end{figure*}

On modern Earth, biological methane production outpaces abiotic production at a rate of about 65:1 \citep{Etiope2013}. Studies have investigated the maximum amount of abiotic methane that might be produced on a terrestrial planet (e.g., see Fig.~\ref{fig:fig6};  \citealp{krissansen2018disequilibrium,wogan2020chemical,Thompson2022case}). These studies find that known abiotic processes are typically unable to produce atmospheres with abundant \ce{CH4 and CO2} due to the redox disequilibrium of these gases, so this pair of gases is critical to observe together when interpreting \ce{CH4} as a biosignature on anoxic worlds. Another gas relevant to interpretation of biogenic methane is CO because CO would be produced by volcanoes on a planet with a highly reducing mantle able to produce abundant volcanic \ce{CH4}. Thus, \ce{CH4} and \ce{CO2} together, with a high \ce{CH4}/\ce{CO} abundance ratio, would be strong evidence pointing towards biogenic methane for anoxic planets. The possibility of methane false positives highlights the need for sufficient NIR capabilities to constrain abundances of \ce{CO2}, \ce{CH4}, and CO.

In addition to the methane flux-producing false positive scenarios above, methane-rich atmospheres from Titan-analogs, post-impact atmospheres, or sub-Neptunes could also be mistaken for an Archean-like biosphere. Possible observational strategies to rule out such false positives are described in Fig.~\ref{fig:fig5}.

\subsection{Other Biosignatures}

Nitrous oxide (\ce{N2O}) is produced by life on Earth through microbial nitrogen metabolisms, which are, in turn, enabled by the presence of abundant oxidants produced by oxygenic photosynthesis. The modern \ce{N2O} atmospheric abundance is low, precluding detection with HWO: 330 ppb (270 ppb pre-industrial). However, past phases of Earth history may have experienced higher amounts of \ce{N2O}. The \ce{N2O} abundance in the Proterozoic atmosphere could have been much greater than today because of the limited ocean oxygenation during that time period, which would have limited trace metals necessary for converting \ce{N2O} to \ce{N2} gas via biological metabolism \citep[e.g.,][]{knowles1982denitrification, pinto2021effects}. A Proterozoic Earth planet orbiting a G-type star can potentially accumulate a \ce{N2O} volume mixing ratio (VMR) of up to 100 ppm (VMR = $10^{-4}$) while a planet around a K-type star has an upper limit of 1000 ppm (VMR = $10^{-3}$) due to the lower UV flux (a major \ce{N2O} sink) from a K vs G star \citep{Schwieterman2022}.

Numerous other biosignatures may be possible (e.g. biogenic sulfur gases; \cite{domagal2011using}) and many others \citep[e.g.,][]{seager2016toward}. Any potential biosignature relevant to HWO must: 1) have absorption features in the HWO wavelength range; 2) accumulate to detectable quantities in a relevant planetary and photochemical context – or have a detectable photochemical byproduct distinguishable from abiotic processes; 3) be studied rigorously to understand potential false positive scenarios. The dominant biosignatures of Earth through time remain the best studied biosignatures for obvious reasons, but we cannot exclude the possibility of other biosignatures on exoplanets. Additional work is needed to mature our understanding of other promising biosignatures, especially given that many have incomplete linelists, incomplete reaction rates, or other deficiencies \citep[e.g.,][]{seager2016toward}. 

\subsection{Corroborating Habitability}
Many of the biosignature false positive scenarios described above could be ruled out by directly confirming surface habitability. Water, while not a biosignature itself, is a critical indicator of planetary habitability. While other solvents have been proposed for life, water has key advantages \citep{pohorille2012water} and it is the most common solvent and polyatomic molecule in the universe.  Indeed, the very definition of the habitable zone is the region around stars where surface liquid water is most probable. Gas-phase water can be detected through its atmospheric spectral features. Direct detection of surface liquid water may be possible on a subset of targets through observation of specular reflection off an ocean (“glint”) \citep[e.g.,][]{robinson2010detecting, lustig2018detecting}, although glint observations will be complicated by the coronagraph inner working angle (IWA) given the necessity of the planet being in or near crescent phase.  

As discussed above, carbon dioxide is a critical part of the planetary context required to interpret methane or organic haze as biosignatures in anoxic atmospheres. Carbon dioxide is also an important greenhouse gas and is detectable on some Earth through time analogs, so constraints on this gas are important for understanding the greenhouse gas budget of a planetary atmosphere.

\section{Physical Parameters}

\begin{figure*}[ht!]
    \centering
    \includegraphics[width=0.75\textwidth]{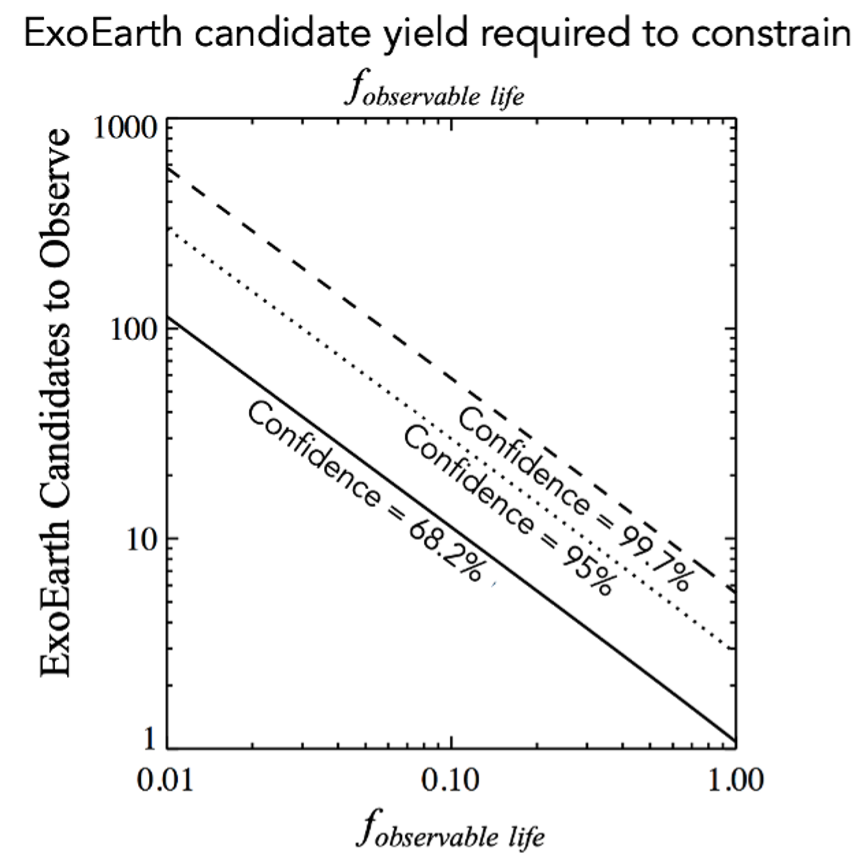}
    \caption{The number of candidate planets that must be observed to constrain various fractions of planets with observable life/biosignatures for given confidence levels. Credit: Christopher Stark.
 }
    \label{fig:fig7}
\end{figure*}

\citetalias{Astro2020} suggests that HWO examine $\sim$25 terrestrial planets in their stars’ habitable zones for signs of habitability and life. This sample size would guarantee seeing at least one planet with biosignatures at 95$\%$ confidence, if the frequency of inhabited planets with observable global biospheres is ~10$\%$ of all candidates. As another example, if, instead, we suppose that 10$\%$ of rocky habitable zone planets represent an Earth twin at a random point in Earth’s evolution, we must observe 33 planets to observe biosignatures at 95$\%$ confidence because 11$\%$ of the “Earths” in the sample will be Hadean-like and uninhabited.  More generally, the number of candidate exoplanets ($N_{ec}$) required to constrain the fraction of planets with a given characteristic x ($\eta_{x}$) at a given confidence level (c) can be written as:
\begin{equation}
    N_{ec}=log(1-c)/log(1-\eta_{x}). 
\end{equation}
We can visualize what this means in Fig.~\ref{fig:fig7}. 

\begin{figure*}[ht!]
    \centering
    \includegraphics[width=0.75\textwidth]{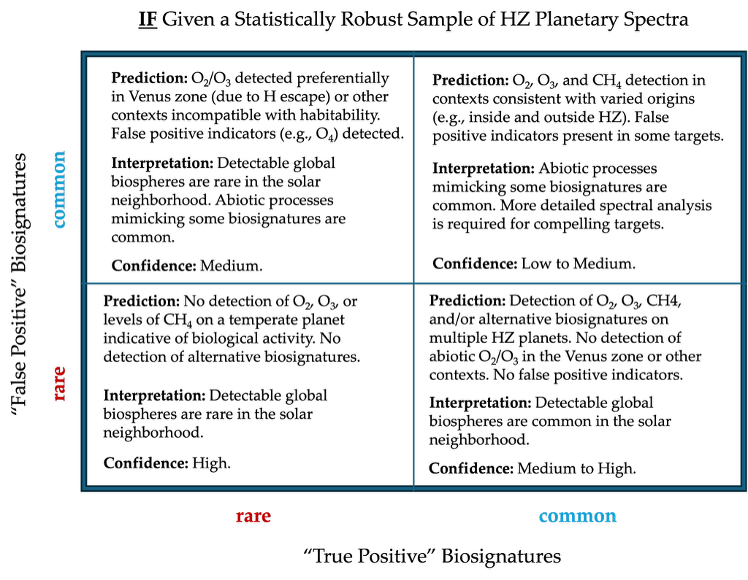}
    \caption{Testable hypotheses about the prevalence of detectable biosignatures and their false positives in the solar neighborhood predicated on how intrinsically common or rare true or false positive biosignatures are. While we often focus on the upper right quadrant, where both biosignatures and their associated false positives are common and individual detections may be ambiguous, other possibilities could be confirmed or refuted even with a capable mission, so long as a statistically robust sample of planets is examined both inside and outside (e.g., interior to) the habitable zone, and false positive disambiguators like \ce{O4} can be detected or excluded. 
 }
    \label{fig:fig8}
\end{figure*}

It is also useful to think about what this means in the case of a null detection. If we examine 25 candidate spectra and do not see signs of life, then we can say that the frequency of habitable planets with observable signs of life is $<$ 10$\%$ of candidate planets in the nearby universe at 95$\%$ confidence, placing the first ever upper limit on the frequency of observable biospheres in the cosmos. Fig.~\ref{fig:fig8} qualitatively summarizes what we might infer about the frequency of planets with various characteristics given different underlying frequencies of true biosignatures and false positives. 

It is also useful to consider how our confidence in interpreting biosignatures may vary based on the spectral type of the parent star. Planets orbiting stars similar to the sun will likely experience star-planet evolutionary histories more akin to Earth around the sun compared to planets orbiting stars with pronounced differences, e.g., dramatically shorter lifetimes, extreme early stellar activity, etc. Our confidence is bolstered in interpreting biosignatures in the atmospheres of planets orbiting sun-like stars. Additionally, more types of biosignature false positives have been identified for planets orbiting later-type stars (e.g., M dwarfs) compared to sun-like stars \citep[e.g.,][]{meadows2017reflections}. We concur with the guidance in \citetalias{Astro2020} to focus the search for habitable planets around sun-like stars as the highest priority targets. However, planets orbiting stars with dramatically different star-planet evolutionary histories compared to the Earth around the sun offer a valuable comparative planetology perspective. These lower mass stars should not be excluded from the target list. We recommend that the target star list place the highest priority on sun-like stars (G and K dwarfs are the highest priority; F dwarfs are second priority given their lower numbers and on-average greater distances to the Earth presenting SNR and IWA issues). More ambitious plans could also include A dwarfs, but given the low numbers of these stars, their shorter lifetimes, and their greater average distances, we recommend that M dwarfs be prioritized over A dwarfs. 

\subsection{Key Physical Parameters}
Below, we describe the information we require from a planet to place important constraints on key properties relevant to the search for habitability and life. 

\subsubsection*{Water (\ce{H2O})}

Detect abundance of atmospheric water vapor to order of magnitude precision for abundances comparable to modern Earth ($\sim$1$\%$ by volume). 

Rationale: Provide context for biosignature interpretation, constraints on planetary greenhouse warming, and information about habitability. 

\subsubsection*{Oxygen (\ce{O2}) 
}

Detect abundance of atmospheric oxygen, a key biosignature, to order of magnitude precision for abundances comparable to Modern Earth’s (21$\%$ by volume).

Rationale: Key biosignature of modern and Proterozoic Earth. 

\subsubsection*{Ozone (\ce{O3}) }

Detect abundance of atmospheric ozone to 1-2 orders of magnitude precision for photochemical abundances consistent with Proterozoic Earth’s oxygen level (i.e., corresponding to 0.1$\%$ present atmospheric level for \ce{O2}).

Rationale: Key biosignature of modern and Proterozoic Earth. 

\subsubsection*{Methane (\ce{CH4})}

Detect abundance of atmospheric methane to order of magnitude precision for abundances comparable to a low-abundance Archean Earth (500 ppm). 

Rationale: Key biosignature for Archean Earth; constraints on greenhouse warming; can rule out oxygen false positives. 

\subsubsection*{Organic Haze }

Detect organic haze if present in anoxic atmospheres for \ce{CH4/CO2} $>$ 0.2.

Rationale: Secondary biosignature of Archean Earth; habitability implications. 

\subsubsection*{Carbon Dioxide (\ce{CO2})}

Detect abundance of atmospheric carbon dioxide to order of magnitude precision for abundances comparable to a low-abundance Archean Earth ($\sim$1$\%$ of the total atmosphere by volume). 

Rationale: Needed for false positive interpretation on anoxic planets; constraints on greenhouse warming. 

\subsubsection*{Carbon Monoxide (CO) }

Place an upper limit on CO if present at $<$1$\%$ or 0.01 bar abundances.

Rationale: Needed for false positive interpretation on anoxic planets.

\subsubsection*{Nitrous Oxide (\ce{N2O}) }

While \ce{N2O} Proterozoic Earth abundances (1-10 ppm) would be challenging to detect, even this amount of \ce{N2O} is suggestive of habitability and life. A more realistic goal is to detect \ce{N2O} at abundances corresponding to an upper limit for a Proterozoic Earth-like planet around a G-star (100 ppm) or K-star (1000 ppm). Minimum requirement: detect \ce{N2O} at $10^{-3}$ near 1.4 $\mu$m. 

Rationale: Secondary biosignature of Proterozoic Earth. 

\subsubsection*{Surface Pressure (or implied \ce{N2})}

Constrain absolute surface pressure to a factor of a few (0.25-0.5 bar).

Rationale: Needed for false positive interpretation on oxygen-rich planets and for inferring planetary climate.\\\\

\begin{figure*}[ht!]
    \centering
    \includegraphics[width=1\textwidth]{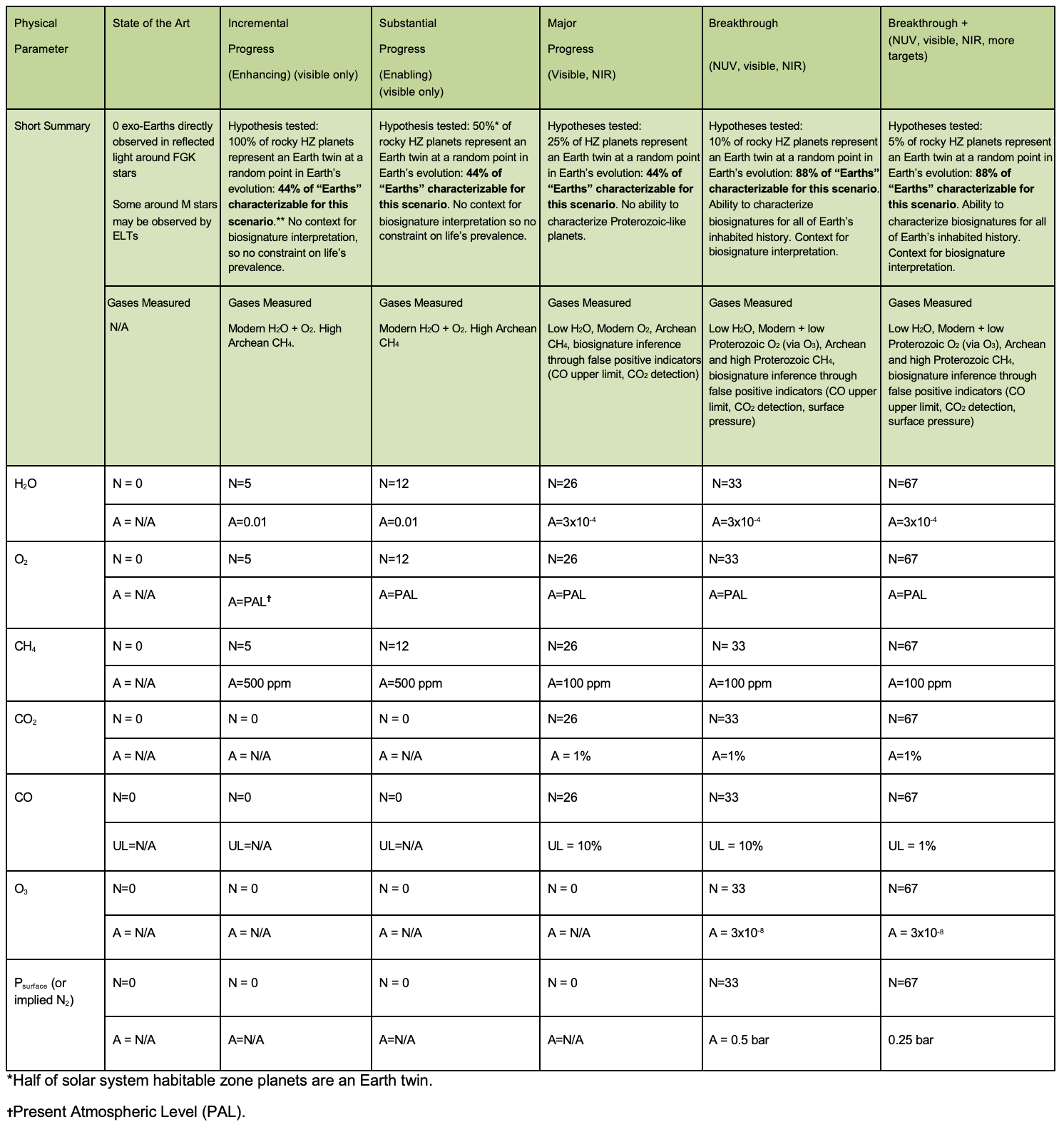}
    \caption{Physical parameters and number of targets required for biosignature search. Columns denote progress in search for life and corresponding constraints on the prevalence of life in the galaxy. Rows denote required physical parameter measurements and number of accessible targets. A denotes required abundance (mixing ratio) constraint, whereas UL denotes required mixing ratio upper limit. N denotes the number of targets. Note that this does not imply that all gases must be precisely constrained for all N targets, merely that the possibility of obtaining constraints via long integrations exists for the most promising targets in the N-sized sample. N is the number of planets that must be observed to rule out each column’s prevalence scenario (X$\%$) to 2 sigma (95$\%$ confidence) based on the fraction of Earth history on which life would have been observable for the scenario.
 }
    \label{fig:fig9}
\end{figure*}

Fig.~\ref{fig:fig9} summarizes the physical parameters that must be measured, along with the requisite abundance or upper limit for gases, and number of targets, in order to search for biosignatures and constrain the prevalence of life in the galaxy. For these calculations, we assume some variable X$\%$ percentage of habitable zone, Earth-sized planets represent Earth at a random point in its history (11$\%$ Hadean, 33$\%$ Archean, 44$\%$ Proterozoic, and 11$\%$ Phanerozoic). Given the atmospheric gases that can be measured, we then calculate the number of targets that must be characterized to rule out an X$\%$ prevalence to 95$\%$ confidence, as shown in Fig.~\ref{fig:fig7}. 

For example, for the “Incremental Progress” case, we assume that every HZ Earth-sized planet is an Earth twin (X=100$\%$) but only Phanerozoic-like \ce{O2} levels and Archean-like \ce{CH4} abundances can be identified given the observatory capabilities. Thus Earth twins with Proterozoic-like biosignatures are invisible, and so only 33$\%$ + 11$\%$ = 44$\%$ of Earth twins can be identified. Consequently, $\log(1-0.95)/\log(1-0.44*100\%) = 5$ targets are needed to rule out this prevalence of life to 95$\%$ confidence. Crucially, however, without the ability to constrain \ce{CO2}, CO, or surface pressure ($N=0$ for each of these parameters), then biosignature false positives cannot be ruled out. We achieve a constraint on the prevalence of Earth-like atmospheric abundances, but not a constraint on the prevalence of life, and so progress is deemed “incremental.” This scenario is also deemed incremental because an implausibly high prevalence of Earth-twins is assumed, $X=100\%$. 

Next, we consider the example of the “Breakthrough Progress” case. Here, we intend to test the more conservative hypothesis that only X=10$\%$ Earth-sized planets in the habitable zone are Earth twins. For the breakthrough case, we assume we can constrain Phanerozoic-like \ce{O2}, Proterozoic-like \ce{O3}, and Archean-like \ce{CH4}. This means that biosignature gases are detectable for $11\% + 44\% + 33\% = 88\%$ of Earth twins. Furthermore, we assume sufficient capabilities to contextualize any potential biosignatures via constraints on \ce{CO, CO2, H2O}, surface pressure. This ability to rule out biosignature false positives for the Earth through time means that $\log(1-0.95)/\log(1-0.88*10\%) = 33$ targets are needed to confirm or rule out a 10$\%$ prevalence of life in the galaxy. The “Substantial Progress” and “Major Progress” scenarios explore in-between cases with different assumed prevalences (X$\%$) and capabilities to identify biosignature false positives.

Note that this SCDD does not discriminate between single and multi-star systems (binaries and higher order). Multi-star systems are present in the current HWO target list, so we expect that most of the science described in this document will be conducted on single as well as binary stars. The presence of multi-star targets is beneficial to this SCDD: they not only increase the expected science yield of the mission, but also likely increase the SNR and spectral coverage of the average planet target. This is because the average FGK star system is closer to us than the average FGK single-star system, which increases the SNR through greater planet photon flux and improves NIR coverage due to larger working angle. Furthermore, the nearest FGK star system to us (Alpha Centauri) is an unusual outlier – it is 2.4 times closer than the next nearest FGK star. This means that, roughly speaking, we expect at least ~2.4 easier working angle and $2.4^2$ ~ 5x more photons from a planet orbiting Alpha Centauri than a similar planet orbiting any other FGK star. If HWO sensitivity to biomarkers is limited by planet flux or inner working angle, then HWO would be sensitive to fainter biomarkers on such a planet. If biomarkers are common but faint, then Alpha Centauri could mean the difference between HWO finding life or not. In addition to the yield and sensitivity advantages, binary stars also allow comparative science between single- and binary- star systems: for example, studies of how binarity affects habitable world formation, evolution, dynamics, etc. (For one example of this, see Elizabeth Newton’s SCDD.)

\section{Description of Observations}

\begin{figure*}[ht!]
    \centering
    \includegraphics[width=0.8\textwidth]{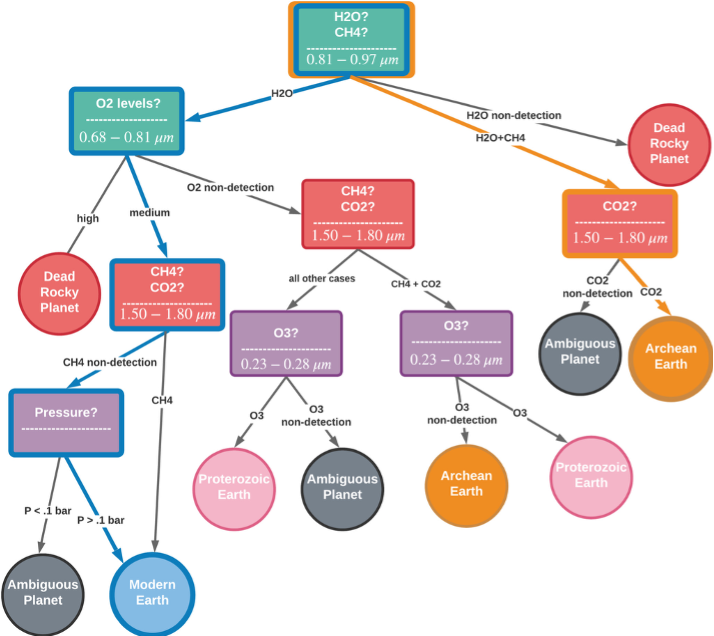}
    \caption{A decision tree strategy for characterizing Earth-like exoplanets \citep{young2024retrievals}. This tree indicates which gasses to observe, and in which order, to detect the biosignatures of Earth through time and rule out relevant false positives. The blue path was followed in spectral retrievals of \citet{young2024retrievals} to categorize modern Earth, and the orange path for Archean Earth. Reproduced with permission from \cite{young2024retrievals}.
 }
    \label{fig:fig10}
\end{figure*}

A wavelength range spanning from $\sim$0.25 – 1.7 $\mu$m provides access to the dominant biosignatures on Earth over its inhabited history according to our best understanding of current geological evidence, in addition to habitability markers and biosignature false positive indicators. Notably, access to these wavelengths also will provide the ability to follow the “decision tree” strategy outlined in \cite{young2024retrievals} for characterizing Earth-like exoplanets (Fig.~\ref{fig:fig10}). This strategy allows observers to classify planets based on phases of Earth history by searching for the dominant biosignatures of Earth through time – and their false positive indicators. Such strategies are especially important to consider given that coronagraphs can only observe spectra within a 10-20$\%$ bandpass at a time. 

\begin{figure*}[ht!]
    \centering
    \includegraphics[width=0.9\textwidth]{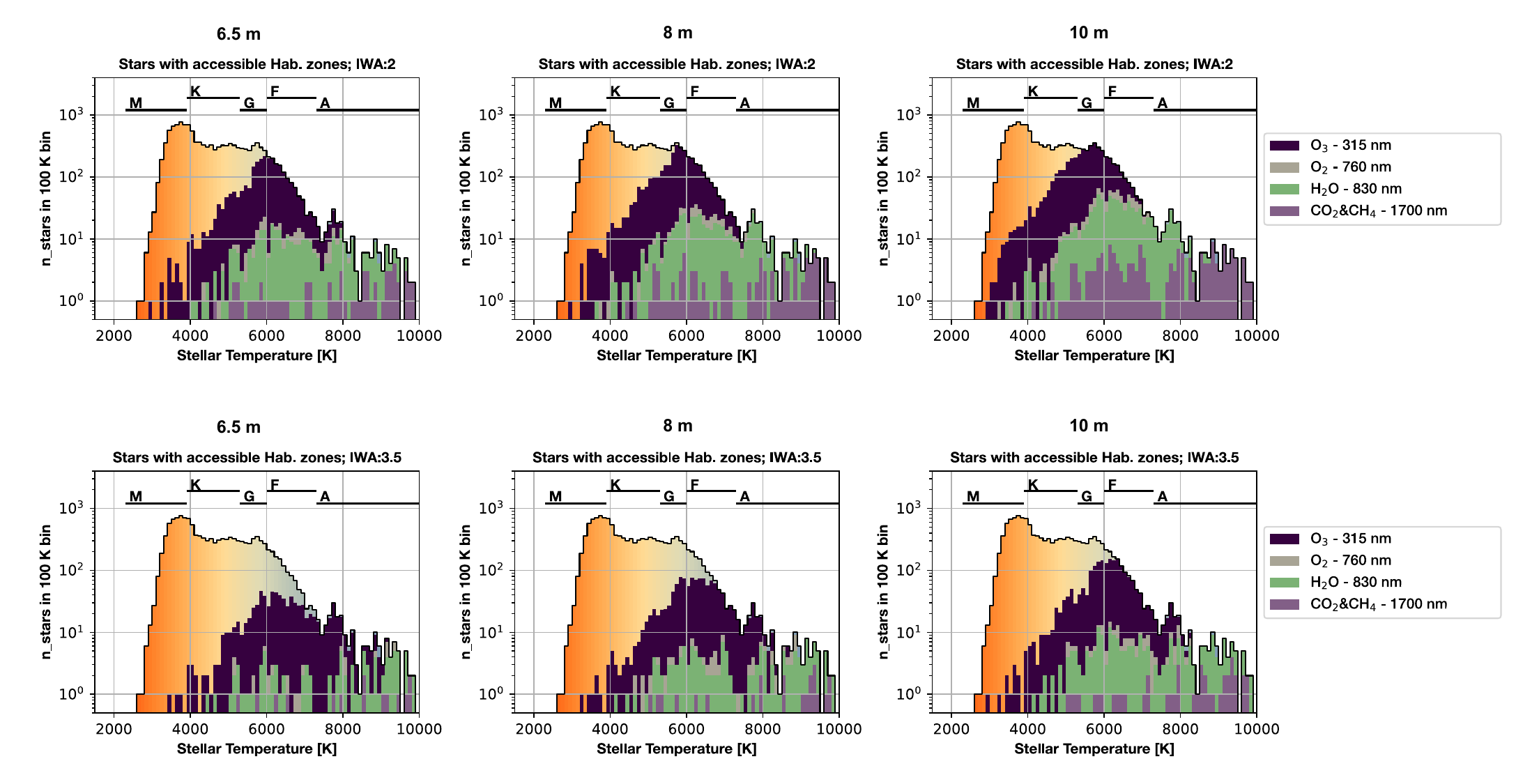}
    \caption{The distribution of potential target stars from the Habitable Worlds Observatory Preliminary Input Catalog (HPIC; \cite{tuchow2024hpic}) as shaded curves. The orange gradient represents all stars in the target volume while the dark purple, gray, green, and light purple colored regions indicate those systems with accessible HZs at the wavelength cutoff given in the legend. Overlain on this are the stars for which various wavelengths are accessible for an IWA of 2$\lambda$/D (top) and 3.5$\lambda$/D (bottom) for an orbiting planet with Earth’s equivalent equilibrium temperature for telescopes with 6.5, 8, and 10 m diameters. Figure and simulations by Vincent Kofman.
 }
    \label{fig:fig11}
\end{figure*}

Of course, it will not be possible to access this entire wavelength range for every HWO target. Fig.~\ref{fig:fig11} shows the distribution of potential target stars from the Habitable Worlds Observatory Preliminary Input Catalog (HPIC) \citep{tuchow2024hpic} as shaded curves. Overlain on this are the stars for which various wavelengths are theoretically accessible for an IWA of 2$\lambda$/D and 3.5$\lambda$/D for a planet with Earth’s equivalent equilibrium temperature around every star for telescopes with 6.5, 8, and 10 m diameters. Unsurprisingly, smaller IWA values and larger telescope apertures provide access to more targets and a wider wavelength range. Ozone in the NUV is the most readily accessible biosignature for the most number of targets based on the parameters considered here. In practice, the number of stars that can actually be accessed for a given architecture will be considerably smaller when exposure time, etc is accounted for. 

\cite{morgan2024hwo} presented HWO yield sensitivities in the NIR and NUV and showed the sensitivity of the NIR yield, in particular, to the IWA and aperture. Of note, the paper concludes that “The NIR achieves 25 exo-Earths characterized only at 9-m aperture and 20 mas IWA, which corresponds to 1 $\lambda$/D.” While these are clearly aggressive numbers, our point is that it is very evident that aperture and IWA are key parameters that will drive the yield of planets that are successfully searched for biosignatures and false positive indicators.

Additional rationale for observation requirements is provided below for each key gas. 

\subsubsection*{Water (\ce{H2O})}
\begin{figure*}[ht!]
    \centering
    \includegraphics[width=0.7\textwidth]{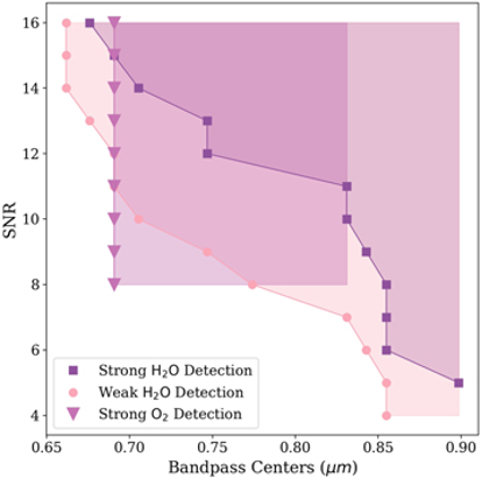}
    \caption{Strength of water vapor and oxygen detection as a function of SNR and spectral bandpass center for a 20$\%$ bandpass. Reproduced with permission from \citet{latouf2023abayesian,latouf2023bbayesian}. 
 }
    \label{fig:fig12}
\end{figure*}

We require observations of water to constrain planetary habitability. SNR = 6 for bandpasses centered between 0.86 and 0.95 is sufficient to provide a strong \ce{H2O} detection for a modern Earth-like water abundance at R = 140 and a 20$\%$ bandpass (Fig.~\ref{fig:fig12}; \cite{latouf2023abayesian,latouf2023bbayesian}). 

\begin{figure*}[ht!]
    \centering
    \includegraphics[width=0.7\textwidth]{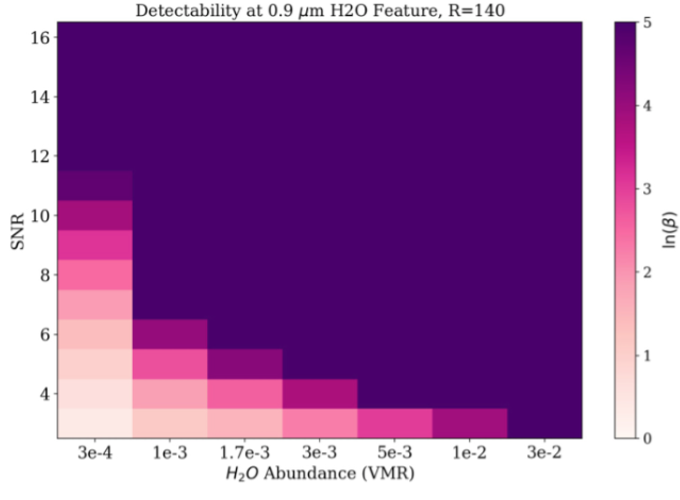}
    \caption{SNR required to detect the 0.9 $\mu$m water vapor feature for varied water abundances. $ln(b) > 5$ represents a “strong” detection according to the criteria used in \cite{latouf2023abayesian,latouf2023bbayesian}. Reproduced with permission from \cite{latouf2023abayesian,latouf2023bbayesian}. 
 }
    \label{fig:fig13}
\end{figure*}

Fig.~\ref{fig:fig13} shows detectability of the water vapor feature near 0.9 $\mu$m for varied water vapor abundances that might represent variability of water on our planet over its history from \citet{latouf2023abayesian}. For the lowest water abundances assumed, SNR = 12 is required to provide a strong detection at R = 140 and 20$\%$ bandpasses.

\subsubsection*{Oxygen (\ce{O2})}
Oxygen’s strongest feature, the “A Band” absorbs at 0.76 $\mu$m, with secondary features at 1.27, 0.69, and 0.63 $\mu$m. As shown in Figure 10, bandpasses centered between 0.7 and 0.83 $\mu$m can achieve a strong detection of \ce{O2} for a spectrum with R = 140 and 20$\%$ bandpasses at SNR $>$ 8, consistent with findings from the LUVOIR final report. The required resolution, which sets the visible channel resolution, is based on sampling the width of the \ce{O2} A-band. 

\subsubsection*{Ozone (\ce{O3})}
Ozone has strong absorption features in the NUV (the Hartley-Huggins bands), and a broad feature at visible wavelengths (the Chappuis band). We consider the Hartley-Huggins NUV band most critical, as it may be the only observable biosignature for a planet similar to mid-Proterozoic Earth; when the Chappuis band is visible, the \ce{O2} A-band will be visible as well. 

There has been concern about the overlap of the NUV ozone band and the NUV sulfur dioxide band (\ce{SO2}) given that both bands’ absorption starts near 0.33 $\mu$m. We investigated the impact of realistic \ce{SO2} in an Earth-like atmosphere on spectral retrievals of ozone using a photochemical model representing Proterozoic Earth with varied \ce{SO2} fluxes. Sulfur dioxide is produced by volcanoes, and a reasonable range for the modern volcanic \ce{SO2} flux has been estimated as $\rm 1-3.5 \times 10^9 ~molecules/cm^2/s$ \citep{zahnle2006loss}. However, we find these fluxes do not produce enough \ce{SO2} to impact the planet’s spectrum. A minor \ce{SO2} feature appears for fluxes of $\rm 9 \times 10^{10} ~molecules/cm^2/s$, which is almost a full an order of magnitude higher than a “high” estimate of the Archean \ce{SO2} flux \citep{zahnle2006loss}, when the planet may have been more volcanically active. 

\begin{figure*}[ht!]
    \centering
    \includegraphics[width=0.85\textwidth]{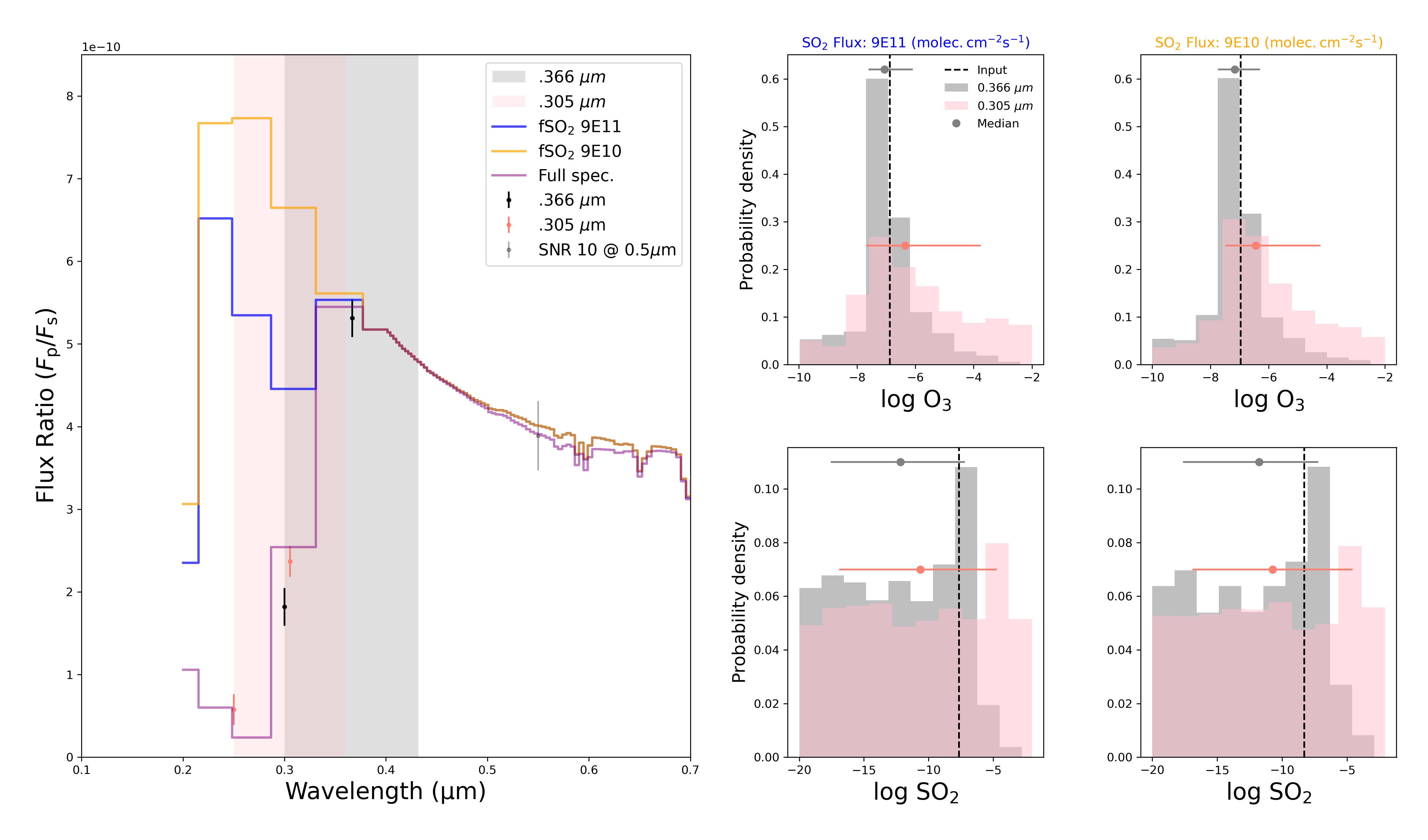}
    \caption{\ce{SO2} is unlikely to confuse ozone retrievals in the NUV. The blue and yellow spectra contain no ozone to clearly show the \ce{SO2} absorption at two \ce{SO2} flux levels indicated in the legend; the purple (full) spectrum includes ozone and \ce{SO2}.The fluxes of \ce{SO2} required to produce an \ce{SO2} spectral signature may be implausibly high in an Earth-like photochemical context given that these fluxes are more than 1-2 orders of magnitude higher than modern Earth's \ce{SO2} flux. Retrievals for two spectral ranges containing two photometric points are included; ozone is constrained by these retrievals, but \ce{SO2} is not. Figure and simulations by Amber Young.
 }
    \label{fig:fig14}
\end{figure*}

 Fig.~\ref{fig:fig14} shows retrievals of \ce{O3} and \ce{SO2} for \ce{SO2} fluxes of $\rm 9 \times 10^{10} ~molecules/cm^2/s$ (weak \ce{SO2} feature, yellow spectrum) and $\rm 9 \times 10^{11} ~molecules/cm^2/s$ (moderate \ce{SO2} feature, blue spectrum), but the latter flux may be implausibly high. Ozone can be retrieved well, but \ce{SO2} remains unconstrained. We emphasize that high abundances of \ce{SO2} may be incompatible with a habitable planet given that high amounts of atmospheric \ce{SO2} such as those required to produce a spectral feature appear to require a planet with little or no surface liquid water \citep{loftus2019sulfate}. High levels of \ce{SO2} may also generate \ce{H2SO4} or \ce{S8} aerosols, depending on the atmospheric redox state \citep[e.g.,][]{hu2013photochemistry}. The highest levels of \ce{SO2} generate \ce{H2SO4} aerosols in our simulations (excluded from spectral retrievals), which would be an independent flag for high \ce{SO2} abundance. None of our atmospheres generate \ce{S8} hazes, which is expected, because \ce{S8} formation requires anoxic conditions \citep[e.g.,][]{pavlov2002mass, zahnle2006loss}.  

\begin{figure*}[ht!]
    \centering
    \includegraphics[width=0.85\textwidth]{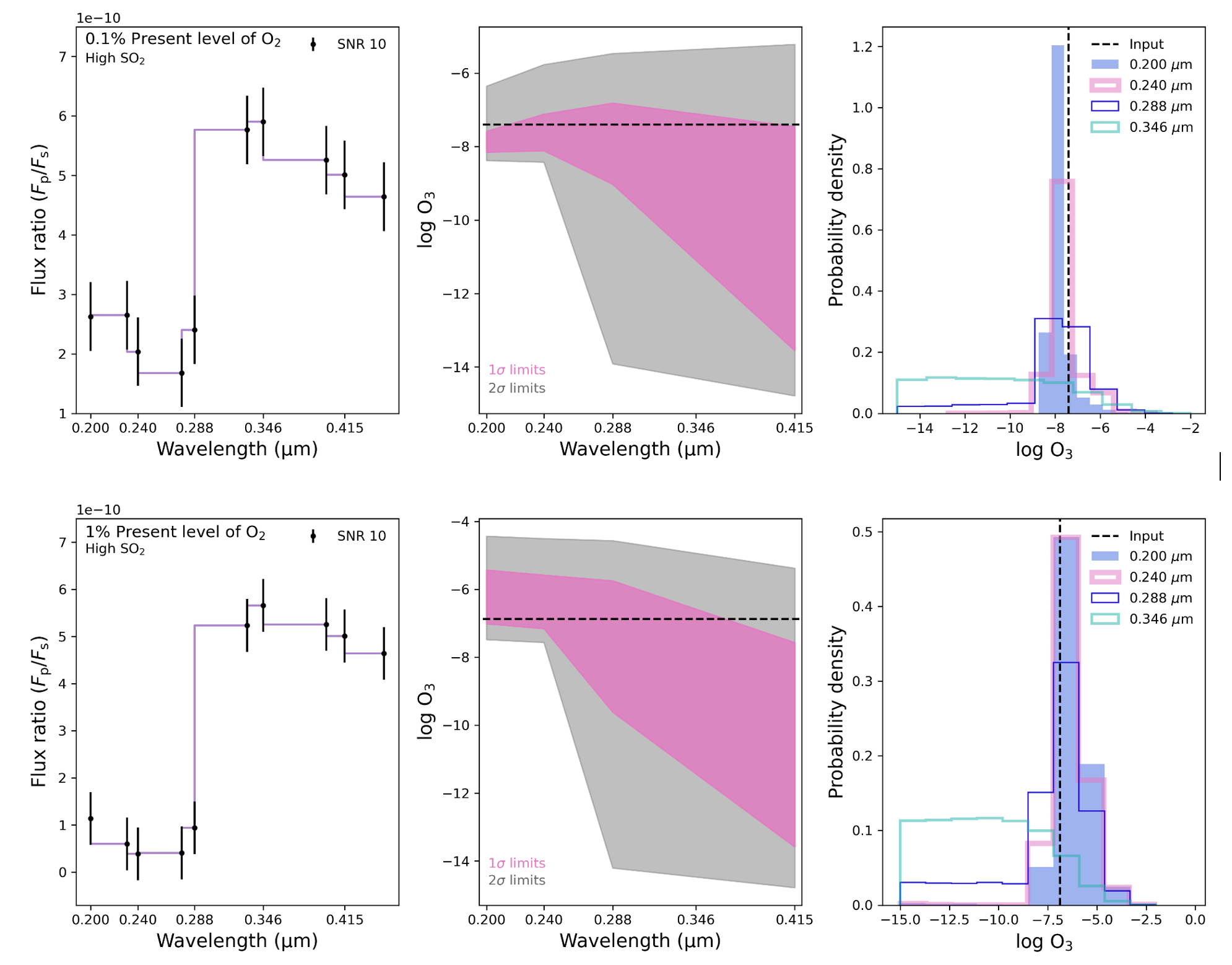}
    \caption{Access to wavelengths $<$ 0.288 $\mu$m is required to adequately constrain ozone for Proterozoic-like planets. Figure and simulations by Amber Young.
 }
    \label{fig:fig15}
\end{figure*}

 Fig.~\ref{fig:fig15} explores the shortest wavelength required to constrain ozone for 0.1$\%$ (top) and 1$\%$ (bottom) the present level of \ce{O2} and “high” \ce{SO2} (a surface flux of $10\times10^{11}$  molecules/cm$^2$/s) at SNR = 10. For this scenario, access to wavelengths $>$ 0.288 $\mu$m is required to adequately constrain Proterozoic ozone. Independently, \cite{damiano2023reflected} recommended a short wavelength cutoff of 0.25 $\mu$m for the best constraints on \ce{O3}.

\subsection*{The short wavelength cutoff}
\begin{figure*}[ht!]
    \centering
    \includegraphics[width=1\textwidth]{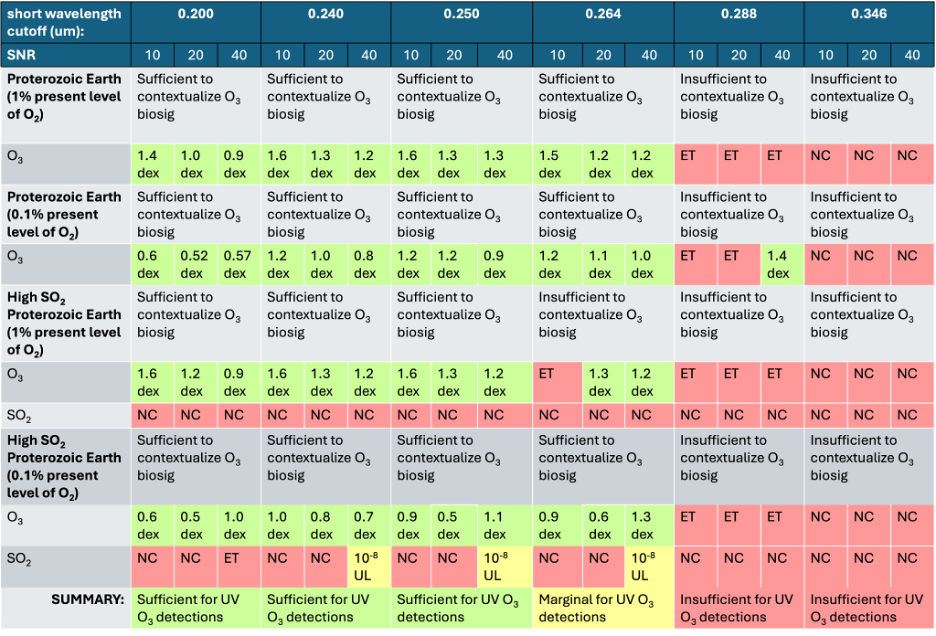}
    \caption{Summary of simulated retrievals showing the extent to which different short wavelength cutoffs enable characterization of Earth-through-time biosignatures, specifically \ce{O3} and its potential spectral contaminant \ce{SO2}. Columns denote shortwave cutoffs and assumed SNRs (R=7), whereas rows represent different atmospheric compositions assumed for the Earth through time (abundances in parentheses). Each grid cell denotes the abundance constraint from a reflected light retrieval using the rfast spectral retrieval model \citep{robinson2023exploring}.  Green grid cells show detections with abundance constraints, yellow grid cells represent upper limits that are useful for contextualizing biosignatures or ruling out known false positive scenarios, whereas red grid cells represent either no constraints (NC) or upper limits with extended tails (ET) that provide no useful contextual information for ruling out biosignature false negatives or assessing habitability. We find a short wavelength cutoff of $\sim$0.25 $\mu$m is necessary for contextualizing Earth-like biosignatures throughout Earth’s evolution. Results based on simulations by Amber Young.
 }
    \label{fig:fig16}
\end{figure*}

Fig.~\ref{fig:fig16} summarizes the results of our team’s spectral retrieval analyses for the short wavelength coronagraph cutoff. Specifically, we investigate the short wavelength cutoff required to contextualize Earth-through-time biosignatures and rule out known false positives for the case of Proterozoic Earth. 

Based on these results, we conclude a short wavelength cutoff of at least 0.25 $\mu$m is needed to constrain \ce{O3} sufficiently well to confidently detect life on Earth at any time in its history; a more restricted wavelength range would potentially lead to ambiguous findings as known biosignature false negatives given that the NUV \ce{O3} feature may have been the only detectable biosignature in the NUV-NIR for a billion years of Earth’s history.

\subsubsection*{Nitrous Oxide (\ce{N2O})}
\cite{tokadjian2024detectability} show that with spectral resolution 7/140/70 in the UV/O/IR and SNR = 20, \ce{N2O} at an atmospheric fraction of $10^{-3}$ is reliably constrained for long wavelength cutoff at least 1.4 $\mu$m. However, \ce{N2O} at an atmospheric fraction of $10^{-4}$ is not detectable even for the full wavelength range (up to 1.8 $\mu$m). Given that K dwarfs accumulate more \ce{N2O} due to their lower UV flux \citep{Schwieterman2022}, our best option for detecting \ce{N2O} on an Earth-like exoplanet through reflection spectroscopy is to focus on K-type stellar hosts and to use a wavelength range that extends to at least 1.4 $\mu$m.

\subsubsection*{Methane (\ce{CH4})}
For modern Earth, methane absorbs weakly near 1.7 $\mu$m, but observing this feature for a modern Earth twin would require extraordinarily high SNR \citep{young2024retrievals}. Methane’s spectral features are stronger, and extend to shorter wavelengths, for planets akin to Archean Earth with higher abundances of \ce{CH4}. \cite{young2024retrievals} shows that methane can be measured near 0.9 $\mu$m for an Archean-like planet with 1000 ppm \ce{CH4} at SNR = 10. The methane feature near 1.7 $\mu$m provides additional leverage for constraining methane on Archean-like planets, and it remains the most observable methane band for planets with more moderate \ce{CH4} levels (e.g., 500 ppm), so access to this band remains critical. Our retrieval analyses for \ce{CH4} are summarized in  Fig.~\ref{fig:fig17}.

\subsubsection*{Organic Haze}
Wavelengths $<$ 0.5 $\mu$m provide access to organic haze absorption, which is a broad spectral feature. Future work, especially spectral retrieval modeling of hazy planets, is needed to constrain organic haze observation requirements. 

\subsubsection*{Carbon Dioxide (\ce{CO2})}  
\begin{figure*}[htb!]
\centering
\includegraphics[width=1\textwidth]{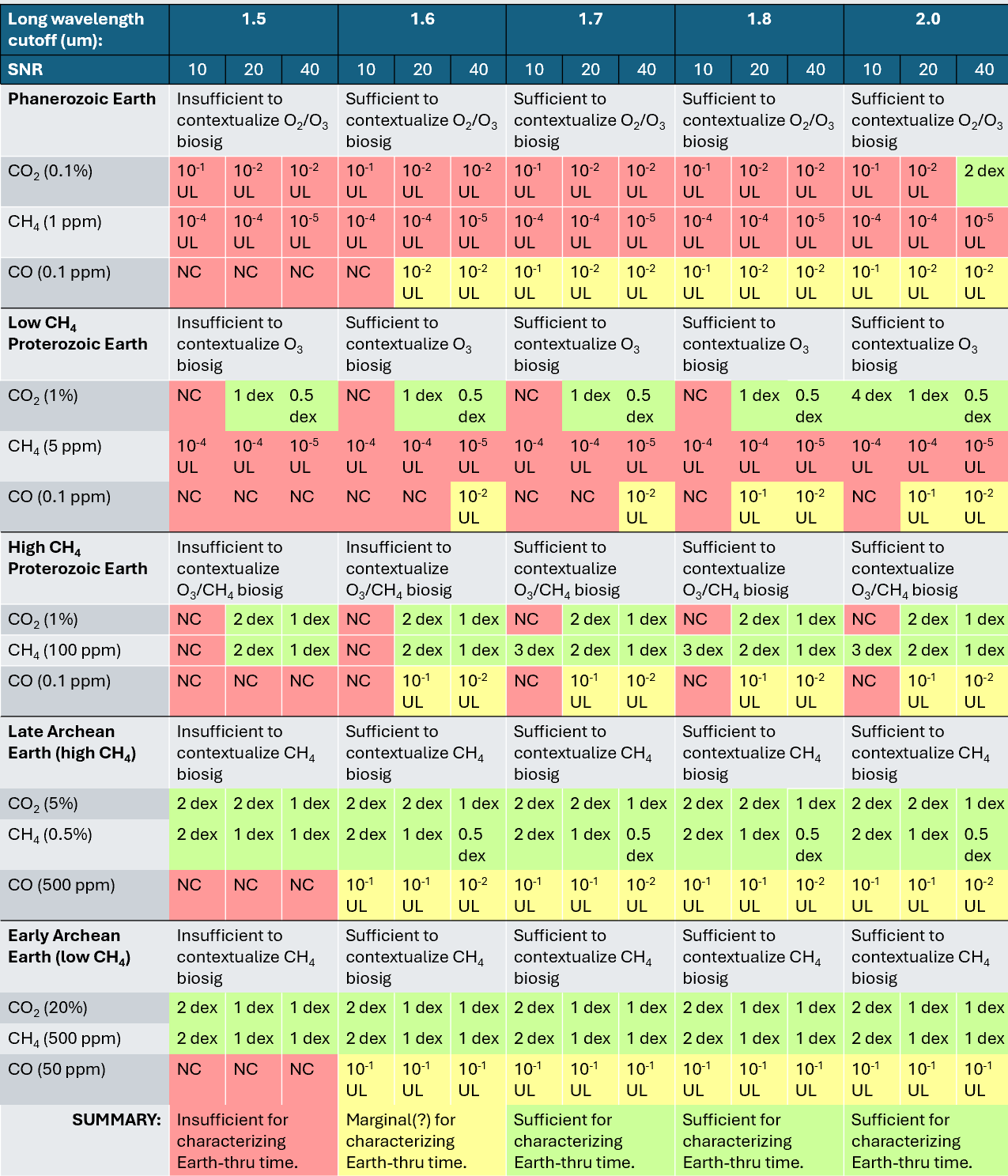}
\caption{Summary of simulated retrievals showing the extent to which different long wavelength cutoffs enable characterization of Earth-through-time biosignatures. Columns denote longwave cutoffs and assumed SNRs (R=7 UV, R=140 VIS, R=70 NIR), whereas rows represent different atmospheric compositions assumed for the Earth through time (abundances in brackets). Each grid cell denotes the abundance constraint from a reflected light retrieval using the rfast spectral retrieval model \citep{robinson2023exploring}.Green grid cells show detections with abundance constraints, yellow grid cells represent upper limits that are useful for contextualizing biosignatures or ruling out known false positive scenarios, whereas red grid cells represent either no constraints (NC) or upper limits (UL) that provide no useful contextual information for ruling out biosignature false positives or assessing habitability. We find a long wavelength cutoff of at least 1.7 $\mu$m is necessary for contextualizing Earth-like biosignatures throughout Earth’s evolution. Results based on simulations by Maxwell Frissell (UW), Anna Grace Ulses (UW), and Samantha Gilbert-Janizek (UW).}
\label{fig:fig17}
\end{figure*}

Modern Earth-like abundances of \ce{CO2} are extremely difficult to constrain (Fig.~\ref{fig:fig17}; \citealp{young2024retrievals}).  Fig.~\ref{fig:fig17} shows that Phanerozoic-like \ce{CO2} levels can be constrained to within 2 orders of magnitude if the long wavelength cutoff can be extended to 2 $\mu$m, but this requires SNR = 40 and would require cooling the telescope. 

For a planet with higher levels of \ce{CO2} (e.g., Archean Earth, with possibly $\sim$1$\%$ atmospheric \ce{CO2} or more), SNR = 10 can place modest constraints on \ce{CO2}, and SNR = 20 can place stronger constraints \citep{young2024retrievals} via the \ce{CO2} features near 1.6 $\mu$m.  Fig.~\ref{fig:fig17} shows that 1$\%$ \ce{CO2} requires high SNR (40) to constrain to within 1 order of magnitude, but \ce{CO2} can be constrained to within 2 orders of magnitude for SNR = 20. Even 20$\%$ \ce{CO2} requires SNR = 20 to constrain to 1 order of magnitude, but it can be constrained to within 2 orders of magnitude for SNR = 10 for a long wavelength cutoff of 1.7 $\mu$m.

\subsubsection*{Carbon Monoxide (CO)} 
Carbon monoxide has absorption features at 1.58 and 2.34 $\mu$m. The 1.58 $\mu$m feature is weak and partially overlaps an absorption feature from carbon dioxide, making spectral retrievals challenging. Fig.~\ref{fig:fig17}  shows that our retrievals were unable to constrain CO in all cases tested; however, a $10^{-1}$ upper limit is possible for a long wavelength cutoff $>$ 1.6 $\mu$m at CO = 50-500 ppm for SNR $>$ 10. 

\subsection*{The long wavelength cutoff}
Fig.~\ref{fig:fig17} summarizes the results of spectral retrieval analyses for the long wavelength coronagraph cutoff. Specifically, we investigate the long wavelength cutoff required to contextualize Earth-through-time biosignatures and rule out known false positives (see above). For the Phanerozoic Earth, we find that detecting atmospheric \ce{CO2}, a potential habitability indicator, would be challenging for any plausible long wavelength cutoff, as would detecting trace amounts of methane. However, access to $>$ 1.6 $\mu$m would enable CO-dominated atmospheres to be excluded, thereby ruling out photochemical oxygen false positives. For the Proterozoic Earth, biogenic methane may be detectable if abundances were at the high end of literature estimates ($\sim$100 ppm), and access to $>$ 1.7 $\mu$m enables biogenic methane detection even if the achievable SNR in the NIR is poor ($\sim$10). This is particularly desirable given the opportunity to simultaneously detect \ce{O3} and \ce{CH4} under this high \ce{CH4} scenario, which would be an especially compelling biosignature with no known false positives. Moreover, access to $>$ 1.6 $\mu$m enables CO-dominated atmospheres to be excluded, once again ruling out photochemical oxygen false positives. Finally, for the Archean Earth, both \ce{CO2} and biogenic \ce{CH4} are detectable for all long wavelength cutoffs we considered. However, access to $>$ 1.6 $\mu$m is needed to rule out CO-dominated atmospheres, and thereby disfavor methane biosignature false positives produced by a reducing planetary interior.

Based on these results, we conclude a long wavelength cutoff of at least 1.7 $\mu$m is needed to constrain \ce{CO2}, \ce{CH4}, and CO sufficiently well to confidently detect life on Earth at any time in its history; a more restricted wavelength range would potentially lead to ambiguous findings as known biosignature false positives could not be excluded. Extending the spectrometer to longer wavelengths ($>$1.7 $\mu$m) to measure additional species not considered here would provide additional leverage to characterize unknown exoplanets on planets observable at those wavelengths (i.e., brighter targets), but we do not recommend that this drive observatory thermal design given the challenges associated with a cold telescope.

\subsubsection*{Other Planetary Properties}
\begin{figure*}[htb!]
    \centering
    \includegraphics[width=0.7\textwidth]{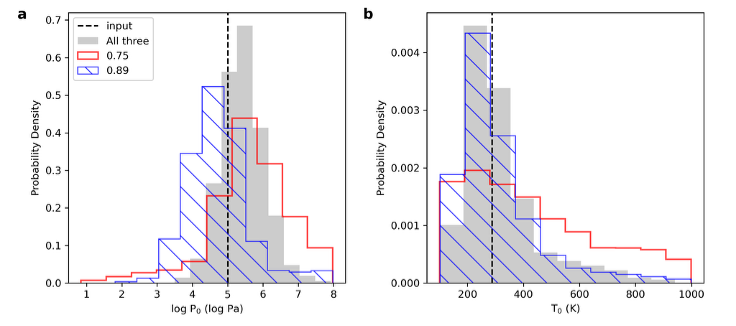}
    \caption{Posterior distributions for global surface pressure (left) and global temperature (right) from simulated SNR = 10 observations. Bandpass centers are indicated in the legend, and the “all three” (gray) retrieval combines information from all three bandpasses considered, which are centered on \ce{O2}, \ce{H2O}, and \ce{CH4} features. Reproduced with permission from \cite{young2024retrievals}.
 }
    \label{fig:fig18}
\end{figure*}

Surface temperature and atmospheric pressure are key planetary properties relevant to habitability and planetary context required to rule our certain false positives such as an oxygen false positive scenario driven by low atmospheric pressures ($\sim$0.1 bar; \cite{wordsworthpierrehumbert2014}). These parameters can be constrained by SNR = 10 observations in 20$\%$ bandpasses in the same observations required to observe \ce{O2} at 0.76 $\mu$m, \ce{H2O} at 0.9 $\mu$m, and \ce{CH4} at 1.65 $\mu$m (assuming R = 140 in the visible and R = 70 in the NIR), possibly allowing us to derive this critical information “for free” during other observations searching for biosignatures and signs of habitability \citep{young2024retrievals}. Fig.~\ref{fig:fig18} shows retrievals of atmospheric pressure and surface temperature from \cite{young2024retrievals}. The temperature information is retrieved from temperature-dependency of opacities, mainly \ce{H2O}, while the pressure information comes from pressure-dependency of opacities. 

\begin{figure*}[htb!]
\centering
\includegraphics[width=1\textwidth]{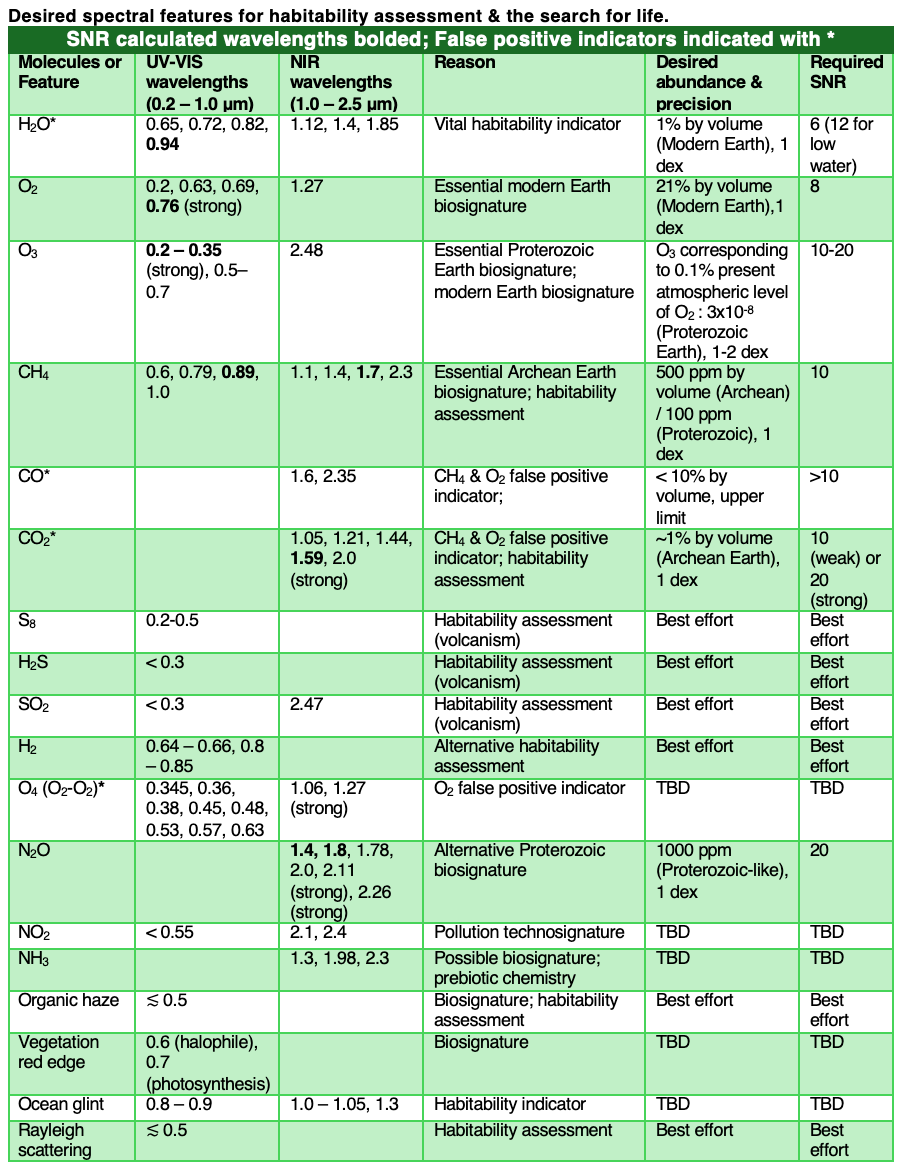}
\caption{Desired spectral features for habitability assessment and search for life as well as calculated requirements for measurements. SNR calculated wavelengths bolded, false positive indicators indicated with *.
\label{fig:fig19}}
\end{figure*}

\begin{figure*}[ht!]
\centering
\includegraphics[width=1\textwidth]{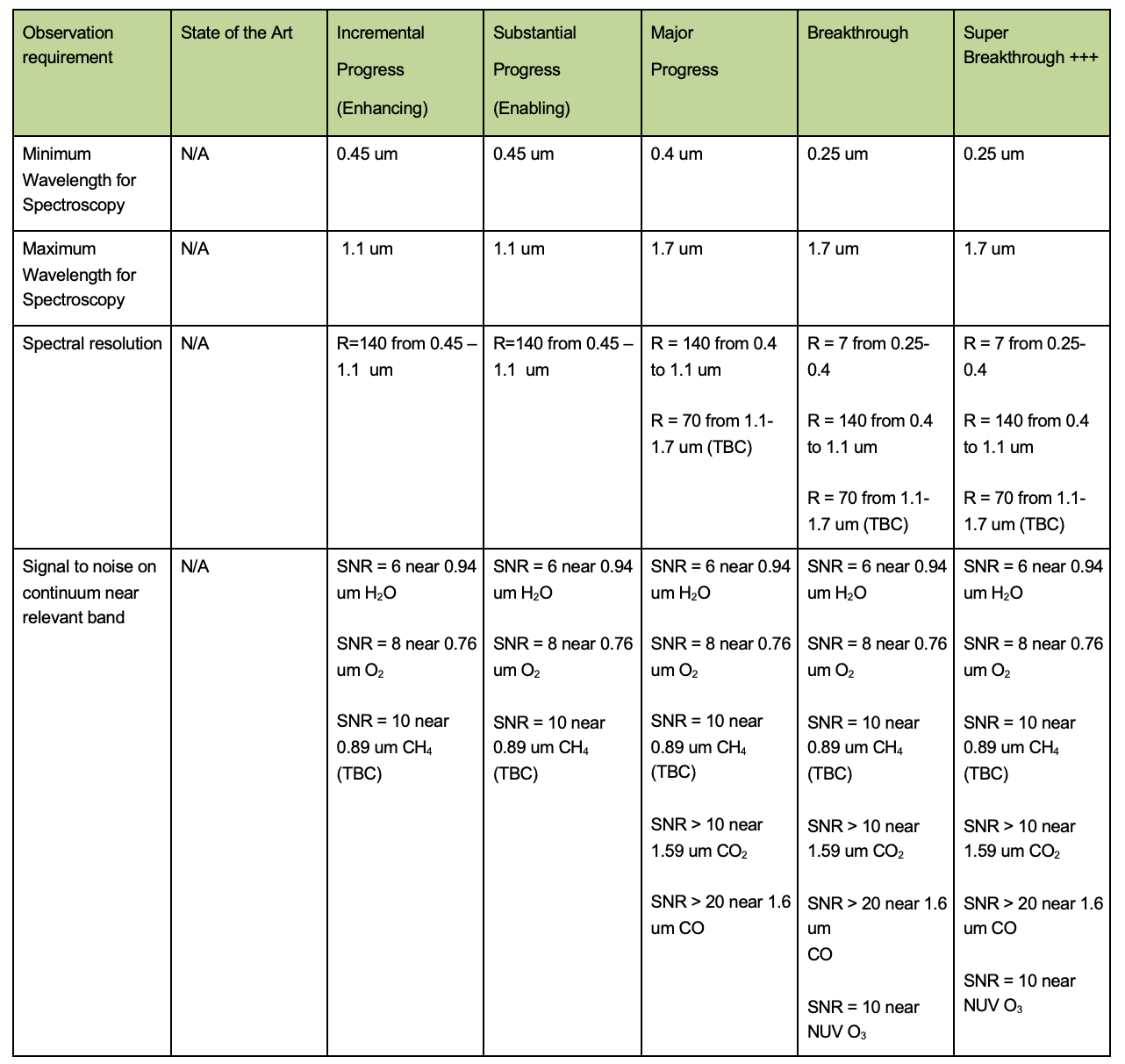}
\caption{Observational requirements to observe living worlds.
\label{fig:fig20}}
\end{figure*}

Fig.~\ref{fig:fig19} and Fig.~\ref{fig:fig20} summarize the requirements for detecting atmospheric gases. Additional work, beyond the scope of this activity, will be required to fill in table cells labeled “TBD,” but these are included here to guide future work. The given wavelengths are near the centers of the absorption bands. Each feature has a finite width, which will depend on concentration (column densities) and other factors. To adequately retrieve the gas in question, it is important to capture both sides of the continuum near the band (e.g., a buffer of $\sim$0.1-0.15 $\mu$m in the NIR). For example, to capture the 1.6 $\mu$m \ce{CO2} band, it is important to extend the spectrum to 1.7 $\mu$m. To capture the 1.65 $\mu$m \ce{CH4} band, it is useful to extend the spectrum 1.8 $\mu$m.

\section{Concluding Remarks}
The search for life beyond the solar system is the most ambitious goal of any NASA mission ever. Success in this endeavor would be one of the most significant discoveries in human history. The astrobiology community has provided a clear map of how to conduct this search in a way that maximizes our likelihood of success. We now await the tool that can conduct this search rigorously. 

\bibliography{author.bib}

@BOOK{Astro2020,
       author = {{National Academies of Sciences, Engineering, and Medicine}},
        title = "{Pathways to Discovery in Astronomy and Astrophysics for the 2020s}",
         year = 2021,
      publisher = {The National Academies Press},
          doi = {10.17226/26141},
       adsurl = {https://ui.adsabs.harvard.edu/abs/2021pdaa.book.....N}
}

@article{young2024retrievals,
  title={Retrievals Applied to a Decision Tree Framework Can Characterize Earthlike Exoplanet Analogs},
  author={Young, Amber V and Crouse, Jaime and Arney, Giada and Domagal-Goldman, Shawn and Robinson, Tyler D and Bastelberger, Sandra T},
  journal={The Planetary Science Journal},
  volume={5},
  number={1},
  pages={7},
  year={2024},
  publisher={IOP Publishing}
}

@article{Meadows2022,
  title={Community report from the biosignatures standards of evidence workshop},
  author={Meadows, Victoria and Graham, Heather and Abrahamsson, Victor and Adam, Zach and Amador-French, Elena and Arney, Giada and Barge, Laurie and Barlow, Erica and Berea, Anamaria and Bose, Maitrayee and others},
  journal={arXiv preprint arXiv:2210.14293},
  year={2022}
}

@article{krissansen2016detecting,
  title={On detecting biospheres from chemical thermodynamic disequilibrium in planetary atmospheres},
  author={Krissansen-Totton, Joshua and Bergsman, David S and Catling, David C},
  journal={Astrobiology},
  volume={16},
  number={1},
  pages={39--67},
  year={2016},
  publisher={Mary Ann Liebert, Inc. 140 Huguenot Street, 3rd Floor New Rochelle, NY 10801 USA}
}

@article{wogan2020chemical,
  title={When is chemical disequilibrium in Earth-like planetary atmospheres a biosignature versus an anti-biosignature? Disequilibria from dead to living worlds},
  author={Wogan, Nicholas F and Catling, David C},
  journal={The Astrophysical Journal},
  volume={892},
  number={2},
  pages={127},
  year={2020},
  publisher={IOP Publishing}
}

@article{sagan1993search,
  title={A search for life on Earth from the Galileo spacecraft},
  author={Sagan, Carl and Thompson, W Reid and Carlson, Robert and Gurnett, Donald and Hord, Charles},
  journal={Nature},
  volume={365},
  number={6448},
  pages={715--721},
  year={1993},
  publisher={Nature Publishing Group UK London}
}

@article{weiss2016physiology,
  title={The physiology and habitat of the last universal common ancestor},
  author={Weiss, Madeline C and Sousa, Filipa L and Mrnjavac, Natalia and Neukirchen, Sinje and Roettger, Mayo and Nelson-Sathi, Shijulal and Martin, William F},
  journal={Nature microbiology},
  volume={1},
  number={9},
  pages={1--8},
  year={2016},
  publisher={Nature Publishing Group}
}

@article{wolfe2018horizontal,
  title={Horizontal gene transfer constrains the timing of methanogen evolution},
  author={Wolfe, Joanna M and Fournier, Gregory P},
  journal={Nature ecology \& evolution},
  volume={2},
  number={5},
  pages={897--903},
  year={2018},
  publisher={Nature Publishing Group UK London}
}

@article{seager2012astrophysical,
  title={An astrophysical view of Earth-based metabolic biosignature gases},
  author={Seager, Sara and Schrenk, Matthew and Bains, William},
  journal={Astrobiology},
  volume={12},
  number={1},
  pages={61--82},
  year={2012},
  publisher={Mary Ann Liebert, Inc. 140 Huguenot Street, 3rd Floor New Rochelle, NY 10801 USA}
}

@article{kaltenegger2017characterize,
  title={How to characterize habitable worlds and signs of life},
  author={Kaltenegger, Lisa},
  journal={Annual Review of Astronomy and Astrophysics},
  volume={55},
  number={1},
  pages={433--485},
  year={2017},
  publisher={Annual Reviews}
}

@article{schwieterman2018exoplanet,
  title={Exoplanet biosignatures: a review of remotely detectable signs of life},
  author={Schwieterman, Edward W and Kiang, Nancy Y and Parenteau, Mary N and Harman, Chester E and DasSarma, Shiladitya and Fisher, Theresa M and Arney, Giada N and Hartnett, Hilairy E and Reinhard, Christopher T and Olson, Stephanie L and others},
  journal={Astrobiology},
  volume={18},
  number={6},
  pages={663--708},
  year={2018},
  publisher={Mary Ann Liebert, Inc. 140 Huguenot Street, 3rd Floor New Rochelle, NY 10801 USA}
}

@article{catlingdavid2018exoplanet,
  title={Exoplanet biosignatures: a framework for their assessment},
  author={Catling, David C and KiangNancy, Y and RobinsonTyler, D and RushbyAndrew, J and others},
  journal={Astrobiology},
  year={2018},
  publisher={Mary Ann Liebert, Inc. 140 Huguenot Street, 3rd Floor New Rochelle, NY 10801 USA}
}

@article{meadows2017reflections,
  title={Reflections on O2 as a biosignature in exoplanetary atmospheres},
  author={Meadows, Victoria S},
  journal={Astrobiology},
  volume={17},
  number={10},
  pages={1022--1052},
  year={2017},
  publisher={Mary Ann Liebert, Inc. 140 Huguenot Street, 3rd Floor New Rochelle, NY 10801 USA}
}

@article{chapman1930,
  title={On ozone and atomic oxygen in the upper atmosphere},
  author={Chapman, Sidney},
  journal={The London, Edinburgh, and Dublin Philosophical Magazine and Journal of Science},
  volume={10},
  number={64},
  pages={369--383},
  year={1930},
  publisher={Taylor \& Francis}
}

@article{segura2003ozone,
  title={Ozone concentrations and ultraviolet fluxes on Earth-like planets around other stars},
  author={Segura, Ant{\'\i}gona and Krelove, Kara and Kasting, James F and Sommerlatt, Darrell and Meadows, Victoria and Crisp, David and Cohen, Martin and Mlawer, Eli},
  journal={Astrobiology},
  volume={3},
  number={4},
  pages={689--708},
  year={2003},
  publisher={Mary Ann Liebert, Inc.}
}

@article{grenfell2014sensitivity,
  title={Sensitivity of biosignatures on Earth-like planets orbiting in the habitable zone of cool M-dwarf Stars to varying stellar UV radiation and surface biomass emissions},
  author={Grenfell, John L and Gebauer, Stefanie and Paris, P v and Godolt, Mareike and Rauer, Heike},
  journal={Planetary and Space Science},
  volume={98},
  pages={66--76},
  year={2014},
  publisher={Elsevier}
}

@article{arney2019k,
  title={The K dwarf advantage for biosignatures on directly imaged exoplanets},
  author={Arney, Giada N},
  journal={The Astrophysical Journal Letters},
  volume={873},
  number={1},
  pages={L7},
  year={2019},
  publisher={IOP Publishing}
}

@article{tabataba2016atmospheric,
  title={Atmospheric effects of stellar cosmic rays on Earth-like exoplanets orbiting M-dwarfs},
  author={Tabataba-Vakili, F and Grenfell, JL and Grie{\ss}meier, J-M and Rauer, H},
  journal={Astronomy \& Astrophysics},
  volume={585},
  pages={A96},
  year={2016},
  publisher={EDP Sciences}
}

@article{tilley2019modeling,
  title={Modeling repeated M dwarf flaring at an Earth-like planet in the habitable zone: atmospheric effects for an unmagnetized planet},
  author={Tilley, Matt A and Segura, Ant{\'\i}gona and Meadows, Victoria and Hawley, Suzanne and Davenport, James},
  journal={Astrobiology},
  volume={19},
  number={1},
  pages={64--86},
  year={2019},
  publisher={Mary Ann Liebert, Inc., publishers 140 Huguenot Street, 3rd Floor New~…}
}

@article{olson2018atmospheric,
  title={Atmospheric seasonality as an exoplanet biosignature},
  author={Olson, Stephanie L and Schwieterman, Edward W and Reinhard, Christopher T and Ridgwell, Andy and Kane, Stephen R and Meadows, Victoria S and Lyons, Timothy W},
  journal={The Astrophysical Journal Letters},
  volume={858},
  number={2},
  pages={L14},
  year={2018},
  publisher={IOP Publishing}
}

@article{schwieterman2016identifying,
  title={Identifying planetary biosignature impostors: spectral features of CO and O4 resulting from abiotic O2/O3 production},
  author={Schwieterman, Edward W and Meadows, Victoria S and Domagal-Goldman, Shawn D and Deming, Drake and Arney, Giada N and Luger, Rodrigo and Harman, Chester E and Misra, Amit and Barnes, Rory},
  journal={The Astrophysical Journal Letters},
  volume={819},
  number={1},
  pages={L13},
  year={2016},
  publisher={IOP Publishing}
}

@ARTICLE{Harrisonetal2005,
       author = {{Harrison}, T.~M. and {Blichert-Toft}, J. and {M{\"u}ller}, W. and {Albarede}, F. and {Holden}, P. and {Mojzsis}, S.~J.},
        title = "{Heterogeneous Hadean Hafnium: Evidence of Continental Crust at 4.4 to 4.5 Ga}",
      journal = {Science},
         year = 2005,
        month = dec,
       volume = {310},
       number = {5756},
        pages = {1947-1950},
          doi = {10.1126/science.1117926},
       adsurl = {https://ui.adsabs.harvard.edu/abs/2005Sci...310.1947H}
}

@ARTICLE{Belletal2015,
       author = {{Bell}, Elizabeth A. and {Boehnke}, Patrick and {Harrison}, T. Mark and {Mao}, Wendy L.},
        title = "{Potentially biogenic carbon preserved in a 4.1 billion-year-old zircon}",
      journal = {Proceedings of the National Academy of Science},
         year = 2015,
        month = nov,
       volume = {112},
       number = {47},
        pages = {14518-14521},
          doi = {10.1073/pnas.1517557112},
       adsurl = {https://ui.adsabs.harvard.edu/abs/2015PNAS..11214518B}
}

@ARTICLE{Nutmanetal2016,
       author = {{Nutman}, Allen P. and {Bennett}, Vickie C. and {Friend}, Clark R.~L. and {van Kranendonk}, Martin J. and {Chivas}, Allan R.},
        title = "{Rapid emergence of life shown by discovery of 3,700-million-year-old microbial structures}",
      journal = {\nat},
         year = 2016,
        month = sep,
       volume = {537},
       number = {7621},
        pages = {535-538},
          doi = {10.1038/nature19355},
       adsurl = {https://ui.adsabs.harvard.edu/abs/2016Natur.537..535N}
}

@article{Zahnleetal2010,
  title={Earth’s earliest atmospheres},
  author={Zahnle, Kevin and Schaefer, Laura and Fegley, Bruce},
  journal={Cold Spring Harbor perspectives in biology},
  volume={2},
  number={10},
  pages={a004895},
  year={2010},
  publisher={Cold Spring Harbor Lab}
}

@ARTICLE{Woganetal2023,
       author = {{Wogan}, Nicholas F. and {Catling}, David C. and {Zahnle}, Kevin J. and {Lupu}, Roxana},
        title = "{Origin-of-life Molecules in the Atmosphere after Big Impacts on the Early Earth}",
      journal = {\psj},
         year = 2023,
        month = sep,
       volume = {4},
       number = {9},
          eid = {169},
        pages = {169},
          doi = {10.3847/PSJ/aced83},
archivePrefix = {arXiv},
       eprint = {2307.09761},
 primaryClass = {astro-ph.EP},
       adsurl = {https://ui.adsabs.harvard.edu/abs/2023PSJ.....4..169W}
}

@ARTICLE{Kasting2014,
       author = {{Kasting}, J.~F.},
        title = "{Modeling the Archean Atmosphere and Climate}",
      journal = {Treatise on Geochemistry},
         year = 2014,
        month = jan,
       volume = {6},
        pages = {157-175},
          doi = {10.1016/b978-0-08-095975-7.01306-1},
       adsurl = {https://ui.adsabs.harvard.edu/abs/2014TrGeo...6..157K}
}

@article{Pavlovetal2000,
  title={Greenhouse warming by CH4 in the atmosphere of early Earth},
  author={Pavlov, Alexander A and Kasting, James F and Brown, Lisa L and Rages, Kathy A and Freedman, Richard},
  journal={Journal of Geophysical Research: Planets},
  volume={105},
  number={E5},
  pages={11981--11990},
  year={2000},
  publisher={Wiley Online Library}
}

@article{WoeseFox1977,
  title={Phylogenetic structure of the prokaryotic domain: the primary kingdoms},
  author={Woese, Carl R and Fox, George E},
  journal={Proceedings of the National Academy of Sciences},
  volume={74},
  number={11},
  pages={5088--5090},
  year={1977},
  publisher={National Academy of Sciences}
}

@article{Uenoetal2006,
  title={Evidence from fluid inclusions for microbial methanogenesis in the early Archaean era},
  author={Ueno, Yuichiro and Yamada, Keita and Yoshida, Naohiro and Maruyama, Shigenori and Isozaki, Yukio},
  journal={Nature},
  volume={440},
  number={7083},
  pages={516--519},
  year={2006},
  publisher={Nature Publishing Group UK London}
}

@article{Thompson2022case,
  title={The case and context for atmospheric methane as an exoplanet biosignature},
  author={Thompson, Maggie A and Krissansen-Totton, Joshua and Wogan, Nicholas and Telus, Myriam and Fortney, Jonathan J},
  journal={Proceedings of the National Academy of Sciences},
  volume={119},
  number={14},
  pages={e2117933119},
  year={2022},
  publisher={National Academy of Sciences}
}

@article{Krissansen2022understanding,
  title={Understanding planetary context to enable life detection on exoplanets and test the Copernican principle},
  author={Krissansen-Totton, Joshua and Thompson, Maggie and Galloway, Max L and Fortney, Jonathan J},
  journal={Nature Astronomy},
  volume={6},
  number={2},
  pages={189--198},
  year={2022},
  publisher={Nature Publishing Group UK London}
}

@article{Schwieterman2019_CO,
  title={Rethinking CO antibiosignatures in the search for life beyond the solar system},
  author={Schwieterman, Edward W and Reinhard, Christopher T and Olson, Stephanie L and Ozaki, Kazumi and Harman, Chester E and Hong, Peng K and Lyons, Timothy W},
  journal={The Astrophysical Journal},
  volume={874},
  number={1},
  pages={9},
  year={2019},
  publisher={IOP Publishing}
}

@article{Trainer2006organic,
  title={Organic haze on Titan and the early Earth},
  author={Trainer, Melissa G and Pavlov, Alexander A and DeWitt, H Langley and Jimenez, Jose L and McKay, Christopher P and Toon, Owen B and Tolbert, Margaret A},
  journal={Proceedings of the National Academy of Sciences},
  volume={103},
  number={48},
  pages={18035--18042},
  year={2006},
  publisher={National Academy of Sciences}
}

@ARTICLE{Arneyetal2016,
       author = {{Arney}, Giada and {Domagal-Goldman}, Shawn D. and {Meadows}, Victoria S. and {Wolf}, Eric T. and {Schwieterman}, Edward and {Charnay}, Benjamin and {Claire}, Mark and {H{\'e}brard}, Eric and {Trainer}, Melissa G.},
        title = "{The Pale Orange Dot: The Spectrum and Habitability of Hazy Archean Earth}",
      journal = {Astrobiology},
         year = 2016,
        month = nov,
       volume = {16},
       number = {11},
        pages = {873-899},
          doi = {10.1089/ast.2015.1422},
archivePrefix = {arXiv},
       eprint = {1610.04515},
 primaryClass = {astro-ph.EP},
       adsurl = {https://ui.adsabs.harvard.edu/abs/2016AsBio..16..873A}
}

@ARTICLE{Arneyetal2017,
       author = {{Arney}, Giada N. and {Meadows}, Victoria S. and {Domagal-Goldman}, Shawn D. and {Deming}, Drake and {Robinson}, Tyler D. and {Tovar}, Guadalupe and {Wolf}, Eric T. and {Schwieterman}, Edward},
        title = "{Pale Orange Dots: The Impact of Organic Haze on the Habitability and Detectability of Earthlike Exoplanets}",
      journal = {\apj},
         year = 2017,
        month = feb,
       volume = {836},
       number = {1},
          eid = {49},
        pages = {49},
          doi = {10.3847/1538-4357/836/1/49},
archivePrefix = {arXiv},
       eprint = {1702.02994},
 primaryClass = {astro-ph.EP},
       adsurl = {https://ui.adsabs.harvard.edu/abs/2017ApJ...836...49A}
}

@ARTICLE{Arneyetal2018,
       author = {{Arney}, Giada and {Domagal-Goldman}, Shawn D. and {Meadows}, Victoria S.},
        title = "{Organic Haze as a Biosignature in Anoxic Earth-like Atmospheres}",
      journal = {Astrobiology},
         year = 2018,
        month = mar,
       volume = {18},
       number = {3},
        pages = {311-329},
          doi = {10.1089/ast.2017.1666},
archivePrefix = {arXiv},
       eprint = {1711.01675},
 primaryClass = {astro-ph.EP},
       adsurl = {https://ui.adsabs.harvard.edu/abs/2018AsBio..18..311A}
}

@article{Zerkle2012bistable,
  title={A bistable organic-rich atmosphere on the Neoarchaean Earth},
  author={Zerkle, Aubrey L and Claire, Mark W and Domagal-Goldman, Shawn D and Farquhar, James and Poulton, Simon W},
  journal={Nature Geoscience},
  volume={5},
  number={5},
  pages={359--363},
  year={2012},
  publisher={Nature Publishing Group UK London}
}

@article{Planavsky2014low,
  title={Low Mid-Proterozoic atmospheric oxygen levels and the delayed rise of animals},
  author={Planavsky, Noah J and Reinhard, Christopher T and Wang, Xiangli and Thomson, Danielle and McGoldrick, Peter and Rainbird, Robert H and Johnson, Thomas and Fischer, Woodward W and Lyons, Timothy W},
  journal={science},
  volume={346},
  number={6209},
  pages={635--638},
  year={2014},
  publisher={American Association for the Advancement of Science}
}

@article{Lyons2014rise,
  title={The rise of oxygen in Earth’s early ocean and atmosphere},
  author={Lyons, Timothy W and Reinhard, Christopher T and Planavsky, Noah J},
  journal={Nature},
  volume={506},
  number={7488},
  pages={307--315},
  year={2014},
  publisher={Nature Publishing Group UK London}
}

@article{Lyons2021oxygenation,
  title={Oxygenation, life, and the planetary system during Earth's middle history: An overview},
  author={Lyons, Timothy W and Diamond, Charles W and Planavsky, Noah J and Reinhard, Christopher T and Li, Chao},
  journal={Astrobiology},
  volume={21},
  number={8},
  pages={906--923},
  year={2021},
  publisher={Mary Ann Liebert, Inc., publishers 140 Huguenot Street, 3rd Floor New~…}
}

@article{Olson2016limited,
  title={Limited role for methane in the mid-Proterozoic greenhouse},
  author={Olson, Stephanie L and Reinhard, Christopher T and Lyons, Timothy W},
  journal={Proceedings of the National Academy of Sciences},
  volume={113},
  number={41},
  pages={11447--11452},
  year={2016},
  publisher={National Academy of Sciences}
}

@article{tokadjian2024detectability,
  title={The Detectability of CH4/CO2/CO and N2O Biosignatures Through Reflection Spectroscopy of Terrestrial Exoplanets},
  author={Tokadjian, Armen and Hu, Renyu and Damiano, Mario},
  journal={The Astronomical Journal},
  volume={168},
  number={6},
  pages={292},
  year={2024},
  publisher={IOP Publishing}
}

@misc{Buick2007,
  title={Did the Proterozoic ‘Canfield Ocean’cause a laughing gas greenhouse?},
  author={Buick, R},
  journal={Geobiology},
  volume={5},
  number={2},
  pages={97--100},
  year={2007},
  publisher={Wiley Online Library}
}

@article{Schwieterman2022,
  title={Evaluating the plausible range of N2O biosignatures on exo-Earths: An integrated biogeochemical, photochemical, and spectral modeling approach},
  author={Schwieterman, Edward W and Olson, Stephanie L and Pidhorodetska, Daria and Reinhard, Christopher T and Ganti, Ainsley and Fauchez, Thomas J and Bastelberger, Sandra T and Crouse, Jaime S and Ridgwell, Andy and Lyons, Timothy W},
  journal={The Astrophysical Journal},
  volume={937},
  number={2},
  pages={109},
  year={2022},
  publisher={IOP Publishing}
}

@article{Reinhard2017,
  title={False negatives for remote life detection on ocean-bearing planets: lessons from the early Earth},
  author={Reinhard, Christopher T and Olson, Stephanie L and Schwieterman, Edward W and Lyons, Timothy W},
  journal={Astrobiology},
  volume={17},
  number={4},
  pages={287--297},
  year={2017},
  publisher={Mary Ann Liebert, Inc. 140 Huguenot Street, 3rd Floor New Rochelle, NY 10801 USA}
}

@article{robinson2018earth,
  title={Earth as an Exoplanet},
  author={Robinson, Tyler D and Reinhard, Christopher T},
  journal={Planetary astrobiology},
  volume={379},
  year={2018}
}

@article{hitchcock1967life,
  title={Life detection by atmospheric analysis},
  author={Hitchcock, Dian R and Lovelock, James E},
  journal={Icarus},
  volume={7},
  number={1-3},
  pages={149--159},
  year={1967},
  publisher={Elsevier}
}

@article{segura2005MDwarfs,
  title={Biosignatures from Earth-like planets around M dwarfs},
  author={Segura, Ant{\'\i}gona and Kasting, James F and Meadows, Victoria and Cohen, Martin and Scalo, John and Crisp, David and Butler, Rebecca AH and Tinetti, Giovanna},
  journal={Astrobiology},
  volume={5},
  number={6},
  pages={706--725},
  year={2005},
  publisher={Mary Ann Liebert, Inc. 2 Madison Avenue Larchmont, NY 10538 USA}
}

@article{wordsworthpierrehumbert2014,
  title={Abiotic oxygen-dominated atmospheres on terrestrial habitable zone planets},
  author={Wordsworth, Robin and Pierrehumbert, Raymond},
  journal={The Astrophysical Journal Letters},
  volume={785},
  number={2},
  pages={L20},
  year={2014},
  publisher={IOP Publishing}
}

@article{Hall2023constraining,
  title={Constraining background N2 inventories on directly imaged terrestrial exoplanets to rule out O2 false positives},
  author={Hall, Sawyer and Krissansen-Totton, Joshua and Robinson, Tyler and Salvador, Arnaud and Fortney, Jonathan J},
  journal={The Astronomical Journal},
  volume={166},
  number={6},
  pages={254},
  year={2023},
  publisher={IOP Publishing}
}

@article{krissansen2021oxygen,
  title={Oxygen false positives on habitable zone planets around sun-like stars},
  author={Krissansen-Totton, Joshua and Fortney, Jonathan J and Nimmo, Francis and Wogan, Nicholas},
  journal={AGU Advances},
  volume={2},
  number={2},
  pages={e2020AV000294},
  year={2021},
  publisher={Wiley Online Library}
}

@article{kopparapu2013habitable,
  title={Habitable zones around main-sequence stars: new estimates},
  author={Kopparapu, Ravi Kumar and Ramirez, Ramses and Kasting, James F and Eymet, Vincent and Robinson, Tyler D and Mahadevan, Suvrath and Terrien, Ryan C and Domagal-Goldman, Shawn and Meadows, Victoria and Deshpande, Rohit},
  journal={The Astrophysical Journal},
  volume={765},
  number={2},
  pages={131},
  year={2013},
  publisher={IOP Publishing}
}

@article{gao2015stability,
  title={Stability of CO2 atmospheres on desiccated M dwarf exoplanets},
  author={Gao, Peter and Hu, Renyu and Robinson, Tyler D and Li, Cheng and Yung, Yuk L},
  journal={The Astrophysical Journal},
  volume={806},
  number={2},
  pages={249},
  year={2015},
  publisher={IOP Publishing}
}

@article{ranjan2023importance,
  title={The importance of the upper atmosphere to CO/O2 runaway on habitable planets orbiting low-mass stars},
  author={Ranjan, Sukrit and Schwieterman, Edward W and Leung, Michaela and Harman, Chester E and Hu, Renyu},
  journal={The Astrophysical Journal Letters},
  volume={958},
  number={1},
  pages={L15},
  year={2023},
  publisher={IOP Publishing}
}

@ARTICLE{Ranjan2020Photochemistry,
       author = {{Ranjan}, Sukrit and {Schwieterman}, Edward W. and {Harman}, Chester and {Fateev}, Alexander and {Sousa-Silva}, Clara and {Seager}, Sara and {Hu}, Renyu},
        title = "{Photochemistry of Anoxic Abiotic Habitable Planet Atmospheres: Impact of New H$_{2}$O Cross Sections}",
      journal = {\apj},
         year = 2020,
        month = jun,
       volume = {896},
       number = {2},
          eid = {148},
        pages = {148},
          doi = {10.3847/1538-4357/ab9363},
archivePrefix = {arXiv},
       eprint = {2004.04185},
 primaryClass = {astro-ph.EP},
       adsurl = {https://ui.adsabs.harvard.edu/abs/2020ApJ...896..148R}
}

@article{Kharecha2005,
  title={A coupled atmosphere--ecosystem model of the early Archean Earth},
  author={Kharecha, P and Kasting, James and Siefert, Janet},
  journal={Geobiology},
  volume={3},
  number={2},
  pages={53--76},
  year={2005},
  publisher={Wiley Online Library}
}

@article{Sauterey2020,
  title={Co-evolution of primitive methane-cycling ecosystems and early Earth’s atmosphere and climate},
  author={Sauterey, Boris and Charnay, Benjamin and Affholder, Antonin and Mazevet, St{\'e}phane and Ferri{\`e}re, R{\'e}gis},
  journal={Nature communications},
  volume={11},
  number={1},
  pages={2705},
  year={2020},
  publisher={Nature Publishing Group UK London}
}

@article{Etiope2013,
  title={Abiotic methane on Earth},
  author={Etiope, Giuseppe and Sherwood Lollar, Barbara},
  journal={Reviews of Geophysics},
  volume={51},
  number={2},
  pages={276--299},
  year={2013},
  publisher={Wiley Online Library}
}

@article{krissansen2018disequilibrium,
  title={Disequilibrium biosignatures over Earth history and implications for detecting exoplanet life},
  author={Krissansen-Totton, Joshua and Olson, Stephanie and Catling, David C},
  journal={Science advances},
  volume={4},
  number={1},
  pages={eaao5747},
  year={2018},
  publisher={American Association for the Advancement of Science}
}

@article{knowles1982denitrification,
  title={Denitrification},
  author={Knowles, Roger},
  journal={Microbiological reviews},
  volume={46},
  number={1},
  pages={43--70},
  year={1982}
}

@article{pinto2021effects,
  title={Effects of dry-wet cycles on nitrous oxide emissions in freshwater sediments: a synthesis},
  author={Pinto, Renata and Weigelhofer, Gabriele and Brito, Ant{\'o}nio Guerreiro and Hein, Thomas},
  journal={PeerJ},
  volume={9},
  pages={e10767},
  year={2021},
  publisher={PeerJ Inc.}
}

@article{domagal2011using,
  title={Using biogenic sulfur gases as remotely detectable biosignatures on anoxic planets},
  author={Domagal-Goldman, Shawn D and Meadows, Victoria S and Claire, Mark W and Kasting, James F},
  journal={Astrobiology},
  volume={11},
  number={5},
  pages={419--441},
  year={2011},
  publisher={Mary Ann Liebert, Inc. 140 Huguenot Street, 3rd Floor New Rochelle, NY 10801 USA}
}

@article{seager2016toward,
  title={Toward a list of molecules as potential biosignature gases for the search for life on exoplanets and applications to terrestrial biochemistry},
  author={Seager, Sara and Bains, W and Petkowski, JJ},
  journal={Astrobiology},
  volume={16},
  number={6},
  pages={465--485},
  year={2016},
  publisher={Mary Ann Liebert, Inc. 140 Huguenot Street, 3rd Floor New Rochelle, NY 10801 USA}
}

@article{pohorille2012water,
  title={Is water the universal solvent for life?},
  author={Pohorille, Andrew and Pratt, Lawrence R},
  journal={Origins of Life and Evolution of Biospheres},
  volume={42},
  number={5},
  pages={405--409},
  year={2012},
  publisher={Springer}
}

@article{robinson2010detecting,
  title={Detecting oceans on extrasolar planets using the glint effect},
  author={Robinson, Tyler D and Meadows, Victoria S and Crisp, David},
  journal={The Astrophysical Journal Letters},
  volume={721},
  number={1},
  pages={L67},
  year={2010},
  publisher={IOP Publishing}
}

@article{lustig2018detecting,
  title={Detecting ocean glint on exoplanets using multiphase mapping},
  author={Lustig-Yaeger, Jacob and Meadows, Victoria S and Mendoza, Guadalupe Tovar and Schwieterman, Edward W and Fujii, Yuka and Luger, Rodrigo and Robinson, Tyler D},
  journal={The Astronomical Journal},
  volume={156},
  number={6},
  pages={301},
  year={2018},
  publisher={IOP Publishing}
}

@article{tuchow2024hpic,
  title={HPIC: the habitable worlds observatory preliminary input catalog},
  author={Tuchow, Noah W and Stark, Christopher C and Mamajek, Eric},
  journal={The Astronomical Journal},
  volume={167},
  number={3},
  pages={139},
  year={2024},
  publisher={IOP Publishing}
}

@inproceedings{morgan2024hwo,
  title={HWO Yield Sensitivities in the NIR and NUV},
  author={Morgan, Rhonda and Savransky, Dmitry and Turmon, Michael and Damiano, Mario and Hu, Renyu and Mennesson, Bertrand and Mamajek, Eric E and Robinson, Tyler D and Tokadjian, Armen},
  booktitle={Space Telescopes and Instrumentation 2024: Optical, Infrared, and Millimeter Wave},
  volume={13092},
  pages={1836--1854},
  year={2024},
  organization={SPIE}
}

@article{latouf2023abayesian,
  title={Bayesian analysis for remote biosignature identification on exoEarths (BARBIE). I. Using grid-based nested sampling in coronagraphy observation simulations for H2O},
  author={Latouf, Natasha and Mandell, Avi M and Villanueva, Geronimo L and Moore, Michael Dane and Susemiehl, Nicholas and Kofman, Vincent and Himes, Michael D},
  journal={The Astronomical Journal},
  volume={166},
  number={3},
  pages={129},
  year={2023},
  publisher={IOP Publishing}
}

@article{latouf2023bbayesian,
  title={Bayesian analysis for remote biosignature identification on exoEarths (BARBIE). II. Using grid-based nested sampling in coronagraphy observation simulations for O2 and O3},
  author={Latouf, Natasha and Mandell, Avi M and Villanueva, Geronimo L and Himes, Michael D and Moore, Michael Dane and Susemiehl, Nicholas and Crouse, Jaime and Domagal-Goldman, Shawn and Arney, Giada and Kofman, Vincent and others},
  journal={The Astronomical Journal},
  volume={167},
  number={1},
  pages={27},
  year={2023},
  publisher={IOP Publishing}
}

@article{zahnle2006loss,
  title={The loss of mass-independent fractionation in sulfur due to a Palaeoproterozoic collapse of atmospheric methane},
  author={Zahnle, K and Claire, Mouterde and Catling, D},
  journal={Geobiology},
  volume={4},
  number={4},
  pages={271--283},
  year={2006},
  publisher={Wiley Online Library}
}

@article{loftus2019sulfate,
  title={Sulfate aerosol hazes and SO2 gas as constraints on rocky exoplanets’ surface liquid water},
  author={Loftus, Kaitlyn and Wordsworth, Robin D and Morley, Caroline V},
  journal={The Astrophysical Journal},
  volume={887},
  number={2},
  pages={231},
  year={2019},
  publisher={IOP Publishing}
}

@article{hu2013photochemistry,
  title={Photochemistry in terrestrial exoplanet atmospheres. II. H2S and SO2 photochemistry in anoxic atmospheres},
  author={Hu, Renyu and Seager, Sara and Bains, William},
  journal={The Astrophysical Journal},
  volume={769},
  number={1},
  pages={6},
  year={2013},
  publisher={IOP Publishing}
}

@article{pavlov2002mass,
  title={Mass-independent fractionation of sulfur isotopes in Archean sediments: strong evidence for an anoxic Archean atmosphere},
  author={Pavlov, AA and Kasting, JF},
  journal={Astrobiology},
  volume={2},
  number={1},
  pages={27--41},
  year={2002},
  publisher={Mary Ann Liebert, Inc.}
}

@article{damiano2023reflected,
  title={Reflected spectroscopy of small exoplanets. III. Probing the UV band to measure biosignature gases},
  author={Damiano, Mario and Hu, Renyu and Mennesson, Bertrand},
  journal={The Astronomical Journal},
  volume={166},
  number={4},
  pages={157},
  year={2023},
  publisher={IOP Publishing}
}

@article{robinson2023exploring,
  title={Exploring and validating exoplanet atmospheric retrievals with solar system analog observations},
  author={Robinson, Tyler D and Salvador, Arnaud},
  journal={The Planetary Science Journal},
  volume={4},
  number={1},
  pages={10},
  year={2023},
  publisher={IOP Publishing}
}

\end{document}